\documentclass[preprint,aps,amsmath,amssymb,12pt]{revtex4}
\usepackage{graphicx}  
\usepackage{float}
\usepackage{bm}        
\usepackage{amssymb}   
\usepackage{slashed}   
\usepackage{amsmath}   
\usepackage{verbatim}  
\usepackage{color} 
\usepackage{xcolor} 
\usepackage{subfig} 
\usepackage{ulem}

\definecolor{Blue}{rgb}{0, 0.1, 0.5}
\graphicspath{{figure/}}
\usepackage{hyperref}

\newcounter{JW}

\begin{document}

\title{Constraints on anomalous quartic gauge couplings via $W\gamma jj$ production at the LHC}

\author{Yu-Chen Guo}
\email{ycguo@lnnu.edu.cn}
\author{Ying-Ying Wang}
\author{Ji-Chong Yang}
\email{yangjichong@lnnu.edu.cn}
\author{Chong-Xing Yue}
\email{cxyue@lnnu.edu.cn}
\affiliation{Department of Physics, Liaoning Normal University, Dalian 116029, China}

\begin{abstract}
The vector boson scattering at the Large Hadron Collider~(LHC) is sensitive to anomalous quartic gauge couplings~(aQGCs).
In this paper, we investigate the aQGC contribution to $ W \gamma jj$ production at the LHC with $\sqrt{s}=13$ TeV in the context of an effective field theory~(EFT).
The unitarity bound is applied as a cut on the energy scale of this production process, which is found to have significant suppressive effects on the signals.
To enhance the statistical significance, we analyse the kinematic and polarization features of the aQGC signals in detail.
We find that the polarization effects induced by the aQGCs are unique and can discriminate the signals from the SM backgrounds well.
With the proposed event selection strategy, we obtain the constraints on the coefficients of dimension-8 operators with current luminosity.
The results indicate that the process $pp \to W \gamma jj$ is powerful for searching for the $O_{M_{2,3,4,5}}$ and $O_{T_{5,6,7}}$ operators.
\end{abstract}

\maketitle

\section{\label{level1}Introduction}

In the past few decades, most of the experimental measurements are in good agreement with the Standard Model~(SM) predictions.
Searching for the new physics beyond SM (BSM) is one of the main goals of current and future colliders.
Among the processes measured at the Large Hadron Collider~(LHC), the vector boson scattering~(VBS) processes provide ideal chances to study BSM.
It is well known that the perturbative unitarity of the longitudinal $W_LZ_L\to W_LZ_L$ scattering is violated without Higgs boson, which sets an upper bound on the mass of the Higgs boson~\cite{VBSandHiggs}.
In other words, with the Higgs boson discovered, the Feynman diagrams of the VBS processes cancel each other and the cross sections do not grow with centre-of-mass (c.m.) energy.
However, such suppression of cross section can be relaxed if there were new physics particles. Consequently, the cross section may be significantly increased and a window to detect BSM is open~\cite{VBSNP}.

A model-independent approach called the SM effective field theory (SMEFT)  \cite{SMEFTReview} has been widely used to search for BSM.
In the SMEFT, the SM is a low energy effective theory of some unknown BSM theory.
When the c.m. energy is not enough to directly produce the new resonance states and when the new physics sector is decoupled, one can integrate out the new physics particles, then the BSM effects become new interactions of known particles.
Formally, the new interactions appear as higher dimensional operators.
The VBS processes are very suitable to search for the existence of new interactions involving electroweak symmetry breaking (EWSB), which is contemplated in many BSM scenarios.
The operators w.r.t. EWSB up to dimension-8 can contribute to the anomalous trilinear gauge couplings~(aTGCs) and anomalous quartic gauge couplings~(aQGCs).
There are many full models that contain these operators, such as anomalous gauge-Higgs couplings, composite Higgs, warped extra dimensions, 2HDM, $U(1)_{L_{\mu}-L_{\tau}}$, as well as axion-like particle scenarios~\cite{AQGCRelatedNP}.

Both aTGCs and aQGCs could have effects on VBS processes~\cite{VBSReview,VBFCut,VBSReview2}.
Unlike aTGCs which also affect the diboson productions and the vector boson fusion~(VBF) processes etc.~\cite{VBSNP,VBSNP3}, the most sensitive processes for aQGCs are the VBS processes.
The dimension-8 operators can contribute to aTGCs and aQGCs independently, therefore, we focus on the dimension-8 anomalous quartic gauge-boson operators.
On the other hand, it is possible that higher dimensional operators contributing to aQGCs exist without dimension-6 operators.
This situation arises in the Born-Infeld (BI) theory proposed in 1934 \cite{BIModel}, which is a nonlinear extension of Maxwell theory motivated by a ``unitarian'' standpoint.
It could provide an upper limit on the strength of the electromagnetic field.
In 1985, the BI theory reborn in models inspired by M-theory \cite{BIplusM}.
We note that the constraint on the BI extension of the SM has recently been presented via dimension-8 operators in the SMEFT~\cite{BIresearch}.

Historically, VBS has been proposed as a means to test the structure of EWSB since the early stage of planning for the Superconducting Super Collider (SSC) \cite{SSC}.
In the past a few years, the study of VBS drew a lot of attention.
The first report of constraints on dimension-8 aQGCs at the LHC is from the same-sign $WW$ production \cite{SameSignWW}.
At present, a number of experimental results in VBS have been obtained, including the electroweak-induced production of $Z\gamma jj$, $W\gamma jj$ at $\sqrt{s}=8\ {\rm TeV}$ and $ZZjj$, $WZjj$, $W^+W^+jj$ at $\sqrt{s}=13\ {\rm TeV}$~\cite{vbsexpall,aw8TeV}.
Theoretical studies are also extensively carried out~\cite{dispersion,{vbsothertheoraitcal}}.
Among these VBS processes, in this paper we consider $W\gamma jj$ production via the scattering between $Z/\gamma$ and $W$ bosons. The next-to-leading order (NLO) QCD corrections to the process $pp\to W\gamma jj$ have been computed in Refs.~\cite{NLOaw,VBFCut}, and the K factor is found to be close to one ($K\approx 0.97$~\cite{VBFCut}).
However the phenomenology of this process with aQGCs needs more exploration.

The SMEFT is only valid under certain energy scale $\Lambda$.
The validity of the SMEFT with dimension-8 operators is an important issue lack of consideration in previous experiments.
The amplitudes of VBS with aQGCs grow as $\mathcal{O}(E^4)$, leading to tree-level unitarity violation at high enough energy~\cite{UnitarityHistory}.
In this case, it is inappropriate to use the SMEFT.
To avoid the violation of unitarity, a unitarity bound should be set.
The unitarity bound is often regarded as a constraint on the coefficient of a high dimensional operator.
However this constraint is not feasible in VBS processes, since the energy scale of the sub-process is not a fixed value but a distribution.
It is proposed that~\cite{shatcut}, to take validity into account, the constraints obtained by experiments should be reported as functions of energy scales.
However, in the $ W\gamma jj$ production, the energy scale of sub-process $\hat{s}=(p_W+p_{\gamma})^2$ is not an observable.
In this work, we find an approximation of $\hat{s}$, based on which the unitarity bounds are applied as cuts on the events at fixed coefficients.
The unitarity bounds will suppress the number of signal events.

To enhance discovery potentiality of the signal, we have to optimize the event selection strategy.
With the approximation of $\hat{s}$, other cuts to cut off the small $\hat{s}$ events become redundant.
Therefore we investigate another important feature of the aQGC contributions, the polarization of the $W$ boson and the resulting angular distribution of the leptons.
The polarization of the $W$ and $Z$ bosons plays an important role in testing the SM~\cite{WPorlarization}.
Angular distribution is a good observable to search for the BSM signals~(an excellent example is the $P_5'$ form factor\cite{P5}) because the differential cross section exposes more information than the total cross section.
While the polarization fractions of the $W$ and $Z$ bosons have been extensively studied within the SM~\cite{WPorlarizationSM,LP}, the angular distribution caused by the polarization effects of aQGCs still need more exploration.

The paper is organized as follows: in section \ref{level2}, we introduce the effective Lagrangian and the corresponding dimension-8 anomalous quartic gauge-boson operators relevant to the $W\gamma jj$ production in VBS processes,
and the experimental constraints on these operators are shown.
In section \ref{level3} we analyse the partial-wave unitarity bounds.
In section \ref{level4}, we firstly propose a cut based on the unitarity bound to ensure that the selected events could be correctly described by the SMEFT.
Then we discuss the kinematic and polarization features of the signal events as well as the event selection strategy.
Based on our event selection strategy, we obtain the constraints on the coefficients of dimension-8 operators with current luminosity at the LHC.
In section \ref{level5}, we present the cross sections and the significance of the aQGC signals in the $\ell\nu\gamma jj$ final state.
Finally, we summarize our results in section \ref{level6}.

\section{\label{level2} Operator basis and constraints from experiments}

The Lagrangian of the SMEFT can be written in terms of an expansion in powers of the inverse of new physics scale $\Lambda$~\cite{SMEFTReview},
\begin{equation}
\begin{split}
&\mathcal{L}_{\rm SMEFT}=\mathcal{L}_{SM}+\sum _i\frac{C_{6i} }{\Lambda^2}\mathcal{O}_{6i}+\sum _j\frac{C_{8j}}{\Lambda^4}\mathcal{O}_{8j}+\ldots,
\end{split}
\label{eq.2.1}
\end{equation}
where $\mathcal{O}_{6i}$ and $\mathcal{O}_{8j}$ are dimension-6 and dimension-8 operators, $C_{6i}/\Lambda ^2$ and $C_{8j}/\Lambda ^4$ are corresponding Wilson coefficients. The effects of BSM are described by higher dimensional operators which are suppressed by $\Lambda$. For one generation fermions, 86 independent operators out of 895 baryon number conserving dimension-8 operators can contribute to QGCs and TGCs~\cite{VBSReview2}.

We list dimension-8 operators affecting aQGCs relevant to $W\gamma jj$ production~\cite{aQGCOperator},
\begin{equation}
\begin{split}
&\mathcal{L}_{aQGC}=\sum _{j} \frac{f_{M_j}}{\Lambda^4}O_{M_j}+\sum _{k} \frac{f_{T_k}}{\Lambda^4}O_{T_k}
\end{split}
\label{eq.2.2}
\end{equation}
with
\begin{equation}
\begin{array}{ll}
O_{M_0}={\rm Tr\left[\widehat{W}_{\mu\nu}\widehat{W}^{\mu\nu}\right]}\times \left[\left(D^{\beta}\Phi \right) ^{\dagger} D^{\beta}\Phi\right],
&O_{M_1}={\rm Tr\left[\widehat{W}_{\mu\nu}\widehat{W}^{\nu\beta}\right]}\times \left[\left(D^{\beta}\Phi \right) ^{\dagger} D^{\mu}\Phi\right],\\
O_{M_2}=\left[B_{\mu\nu}B^{\mu\nu}\right]\times \left[\left(D^{\beta}\Phi \right) ^{\dagger} D^{\beta}\Phi\right],
&O_{M_3}=\left[B_{\mu\nu}B^{\nu\beta}\right]\times \left[\left(D^{\beta}\Phi \right) ^{\dagger} D^{\mu}\Phi\right],\\
O_{M_4}=\left[\left(D_{\mu}\Phi \right)^{\dagger}\widehat{W}_{\beta\nu} D^{\mu}\Phi\right]\times B^{\beta\nu},
&O_{M_5}=\left[\left(D_{\mu}\Phi \right)^{\dagger}\widehat{W}_{\beta\nu} D_{\nu}\Phi\right]\times B^{\beta\mu} + h.c.,\\
O_{M_7}=\left(D_{\mu}\Phi \right)^{\dagger}\widehat{W}_{\beta\nu}\widehat{W}_{\beta\mu} D_{\nu}\Phi,&\\
\end{array}
\label{eq.2.4}
\end{equation}
\begin{equation}
\begin{array}{ll}
O_{T_0}={\rm Tr}\left[\widehat{W}_{\mu\nu}\widehat{W}^{\mu\nu}\right]\times {\rm Tr}\left[\widehat{W}_{\alpha\beta}\widehat{W}^{\alpha\beta}\right],
&O_{T_1}={\rm Tr}\left[\widehat{W}_{\alpha\nu}\widehat{W}^{\mu\beta}\right]\times {\rm Tr}\left[\widehat{W}_{\mu\beta}\widehat{W}^{\alpha\nu}\right],\\
O_{T_2}={\rm Tr}\left[\widehat{W}_{\alpha\mu}\widehat{W}^{\mu\beta}\right]\times {\rm Tr}\left[\widehat{W}_{\beta\nu}\widehat{W}^{\nu\alpha}\right],
&O_{T_5}={\rm Tr}\left[\widehat{W}_{\mu\nu}\widehat{W}^{\mu\nu}\right]\times B_{\alpha\beta}B^{\alpha\beta},\\
O_{T_6}={\rm Tr}\left[\widehat{W}_{\alpha\nu}\widehat{W}^{\mu\beta}\right]\times B_{\mu\beta}B^{\alpha\nu},
&O_{T_7}={\rm Tr}\left[\widehat{W}_{\alpha\mu}\widehat{W}^{\mu\beta}\right]\times B_{\beta\nu}B^{\nu\alpha},\\
\end{array}
\label{eq.2.5}
\end{equation}
where $\widehat{W}\equiv \vec{\sigma}\cdot \vec{W}/2$ with $\sigma$ being the Pauli matrices and $\vec{W}\equiv \{W^1, W^2, W^3\}$.

The tightest constraints on the coefficients of the corresponding operators are obtained via $WWjj$, $WZjj$, $ZZjj$ and $Z\gamma jj$ channels by CMS experiments at $ \sqrt{s}=13$ TeV~\cite{newdatawvzv, {newdataza}}, which are listed in Table~\ref{tab.1}.
\begin{table}
\caption{\label{tab.1}The constraints on the coefficients obtained by CMS experiments.}
\centering
\begin{tabular}{c|c||c|c}
\hline
coefficient & constraint & coefficient & constraint \\
\hline
$f_{M_0}/\Lambda ^4\;({\rm TeV^{-4}})$ & $[-0.69,0.70]$~\cite{newdatawvzv} & $f_{T_0}/\Lambda ^4\;({\rm TeV^{-4}})$ & $[-0.12,0.11]$~\cite{newdatawvzv} \\
$f_{M_1}/\Lambda ^4\;({\rm TeV^{-4}})$ & $[-2.0,2.1]$~\cite{newdatawvzv} & $f_{T_1}/\Lambda ^4\;({\rm TeV^{-4}})$ & $[-0.12,0.13]$~\cite{newdatawvzv} \\
$f_{M_2}/\Lambda ^4\;({\rm TeV^{-4}})$ & $[-8.2,8.0]$~\cite{newdataza} & $f_{T_2}/\Lambda ^4\;({\rm TeV^{-4}})$ & $[-0.28,0.28]$~\cite{newdatawvzv} \\
$f_{M_3}/\Lambda ^4\;({\rm TeV^{-4}})$ & $[-21,21]$~\cite{newdataza} & $f_{T_5}/\Lambda ^4\;({\rm TeV^{-4}})$ & $[-0.7,0.74]$~\cite{newdataza} \\
$f_{M_4}/\Lambda ^4\;({\rm TeV^{-4}})$ & $[-15,16]$~\cite{newdataza} & $f_{T_6}/\Lambda ^4\;({\rm TeV^{-4}})$ & $[-1.6,1.7]$~\cite{newdataza} \\
$f_{M_5}/\Lambda ^4\;({\rm TeV^{-4}})$ & $[-25,24]$~\cite{newdataza} & $f_{T_7}/\Lambda ^4\;({\rm TeV^{-4}})$ & $[-2.6,2.8]$~\cite{newdataza} \\
$f_{M_7}/\Lambda ^4\;({\rm TeV^{-4}})$ & $[-3.4,3.4]$~\cite{newdatawvzv} & & \\
\hline
\end{tabular}
\end{table}

The aQGC vertices relevant to $W\gamma jj$ channel are $W^+W^-\gamma\gamma$ and $ W^+W^-Z\gamma$ vertices, which are
\begin{equation}
\begin{array}{ll}
V_{WWZ\gamma,0}=F^{\mu\alpha}Z_{\mu\beta}(W^+_{\alpha}W^{-\beta}+W^-_{\alpha}W^{+\beta}),
&V_{WWZ\gamma,1}=F^{\mu\alpha}Z_{\alpha}(W^+_{\mu\beta}W^{-\beta}+W^-_{\mu\beta}W^{+\beta}),\\
V_{WWZ\gamma,2}=F^{\mu\nu}Z_{\mu\nu}W^+_{\alpha}W^{-\alpha},
&V_{WWZ\gamma,3}=F^{\mu\alpha}Z^{\beta}(W^+_{\mu\alpha}W^-_{\beta}+W^-_{\mu\alpha}W^+_{\beta}),\\
V_{WWZ\gamma,4}=F^{\mu\alpha}Z^{\beta}(W^+_{\mu\beta}W^-_{\alpha}+W^-_{\mu\beta}W^+_{\alpha}),
&V_{WWZ\gamma,5}=F^{\mu\nu}Z_{\mu\nu}W^{+\alpha\beta}W^-_{\alpha\beta},\\
V_{WWZ\gamma,6}=F^{\mu\alpha}Z_{\mu\beta}(W^+_{\nu\alpha}W^{-\nu\beta}+W^-_{\nu\alpha}W^{+\nu\beta}),
&V_{WWZ\gamma,7}=F^{\mu\nu}Z^{\alpha\beta}(W^+_{\mu\nu}W^-_{\alpha\beta}+W^-_{\mu\nu}W^+_{\alpha\beta}).
\end{array}
\label{eq.2.6}
\end{equation}
\begin{equation}
\begin{array}{ll}
V_{WW\gamma\gamma,0}=F_{\mu\nu}F^{\mu\nu}W^{+\alpha}W^-_{\alpha},
&V_{WW\gamma\gamma,1}=F_{\mu\nu}F^{\mu\alpha}W^{+\nu}W^-_{\alpha},\\
V_{WW\gamma\gamma,2}=F_{\mu\nu}F^{\mu\nu}W^+_{\alpha\beta}W^{-\alpha\beta},
&V_{WW\gamma\gamma,3}=F_{\mu\nu}F^{\nu\alpha}W^+_{\alpha\beta}W^{-\beta\mu},\\
V_{WW\gamma\gamma,4}=F_{\mu\nu}F^{\alpha\beta}W_{\mu\nu}^+W^{-\alpha\beta},& \\
\end{array}
\label{eq.2.7}
\end{equation}
and the coefficients are
\begin{equation}
\begin{split}
&\begin{array}{ll}
\alpha_{WWZ\gamma,0}=\frac{e^2v^2}{8\Lambda ^4}\left(\frac{c_W^2}{s_W^2}f_{M_5}-f_{M_5}-\frac{c_W}{s_W}f_{M_1}+2\frac{c_W}{s_W}f_{M_3}+\frac{c_W}{2s_W}f_{M_7}\right),&\\
\alpha_{WWZ\gamma,1}=\frac{e^2v^2}{8\Lambda ^4}\left(-\frac{1}{2}\left(\frac{c_W}{s_W}+\frac{s_W}{c_W}\right)f_{M_7}-f_{M_5}-\frac{c_W^2}{s_W^2}f_{M_5}\right),&\\
\alpha_{WWZ\gamma,2}=\frac{e^2v^2}{8\Lambda ^4}\left(\frac{c_W^2}{s_W^2}f_{M_4}-f_{M_4}+2\frac{c_W}{s_W}f_{M_0}-4\frac{c_W}{s_W}f_{M_2}\right),&\\
\end{array}\\
&\begin{array}{ll}
\alpha_{WWZ\gamma,3}=\frac{e^2v^2}{8\Lambda ^4}\left(-\frac{c_W^2}{s_W^2}f_{M_4}-f_{M_4}\right),\;
&\alpha_{WWZ\gamma,4}=\frac{e^2v^2}{8\Lambda ^4}\left(\frac{1}{2}\left(\frac{c_W}{s_W}+\frac{s_W}{c_W}\right)f_{M_7}-f_{M_5}-\frac{c_W^2}{s_W^2}f_{M_5}\right),\\
\alpha_{WWZ\gamma,5}=\frac{2c_Ws_W}{\Lambda^4}\left(f_{T_0}-f_{T_5}\right),\;
&\alpha_{WWZ\gamma,6}=\frac{c_Ws_W}{\Lambda^4}\left(f_{T_2}-f_{T_7}\right),\;\\
\alpha_{WWZ\gamma,7}=\frac{c_Ws_W}{\Lambda^4}\left(f_{T_1}-f_{T_6}\right),&
\end{array}
\end{split}
\label{eq.2.8}
\end{equation}
\begin{equation}
\begin{array}{ll}
\alpha_{WW\gamma\gamma,0}=\frac{e^2v^2}{8\Lambda ^4}\left(f_{M_0}+\frac{c_W}{s_W}f_{M_4}+2\frac{c_W^2}{s_W^2}f_{M_2}\right),&\\
\alpha_{WW\gamma\gamma,1}=\frac{e^2v^2}{8\Lambda ^4}\left(\frac{1}{2}f_{M_7}+2\frac{c_W}{s_W}f_{M_5}-f_{M_1}-2\frac{c_W^2}{s_W^2}f_{M_3}\right),
&\alpha_{WW\gamma\gamma,2}=\frac{1}{\Lambda ^4}\left(s_W^2f_{T_0}+c_W^2f_{T_5}\right),\;\\
\alpha_{WW\gamma\gamma,3}=\frac{1}{\Lambda ^4}\left(s_W^2f_{T_2}+c_W^2f_{T_7}\right),\;
&\alpha_{WW\gamma\gamma,4}=\frac{1}{\Lambda ^4}\left(s_W^2f_{T_1}+c_W^2f_{T_6}\right).
\end{array}
\label{eq.2.9}
\end{equation}

Note that the vertices $V_{WWZ\gamma,0,1,2,3,4}$ and $V_{AAWW,0,1}$ are dimension-6 derived from $O_{M_i}$, and the other vertices are dimension-8 derived from $O_{T_i}$.

\section{\label{level3}Unitarity bounds}

Unlike in the SM, the cross section of the VBS process with aQGCs grows with c.m. energy.
Such feature opens a window to detect aQGCs at higher energies. However, the cross section with aQGCs will violate unitarity at certain energy scale.
The unitarity violation indicates that the SMEFT is no longer valid to describe the phenomenon at such high energies perturbatively.

Considering the process $V_{1,\lambda _1}V_{2,\lambda _2}\to V_{3,\lambda _3}V_{4,\lambda _4}$, where $V_i$ are vector bosons, $\lambda _i$ correspond to the helicities of $V_i$, and therefore $\lambda _i=\pm 1$ for photons, and $\lambda _i=\pm 1, 0$ for $W^{\pm},Z$ bosons, its amplitudes can be expanded as~\cite{PartialWave,UnitaryBound}
\begin{equation}
\begin{split}
&\mathcal{M}(V_{1,\lambda _1}W^+_{\lambda _2}\to \gamma_{\lambda _3}W^+_{\lambda _4})=8\pi \sum _{J}\left(2J+1\right)\sqrt{1+\delta _{\lambda _1\lambda _2}}\sqrt{1+\delta _{\lambda _3\lambda _4}}e^{i(\lambda-\lambda ') \varphi}d^J_{\lambda \lambda '}(\theta) T^J,\\
\end{split}
\label{eq.3.1}
\end{equation}
where $V_1$ is $\gamma$ or $Z$ boson, $\lambda = \lambda _1-\lambda _2$, $\lambda ' =\lambda _3-\lambda _4$, $\theta$ and $\phi$ are the zenith and azimuth angles of the $\gamma$ in the final state, $d^J_{\lambda \lambda '}(\theta)$ are the Wigner $d$-functions~\cite{PartialWave}, and $T^J$ are coefficients of the expansion which can be obtained via Eq.~(\ref{eq.3.1}). Partial-wave unitarity for the elastic channels requires $|T^J|\leq 2$~\cite{UnitaryBound}, which is widely used in previous works~\cite{UnitaryBoundOthers}.

\subsection{\label{level3.1}Partial-wave expansions of the \texorpdfstring{$W\gamma\to W\gamma$}{WA to WA} amplitudes}

We calculate the partial-wave expansions of the $W\gamma\to W\gamma$ amplitudes with one dimension-8 operator at a time.
Denoting $\mathcal{M}^{f_{X}}$ as the amplitude with only $O_X$ operator, for $O_{M_{2,3,4,5,7}}$ and $O_{T_{5,6,7}}$, which can be derived by using Eq.~(\ref{eq.2.9}) as
\begin{equation}
\begin{split}
&\mathcal{M}^{f_{M_4}}(W^+\gamma\to W^+\gamma)=\frac{c_W}{s_W}\frac{f_{M_4}}{f_{M_0}}\mathcal{M}^{f_{M_0}}(W^+\gamma\to W^+\gamma),\\
&\mathcal{M}^{f_{M_2}}(W^+\gamma\to W^+\gamma)=\frac{2c_W^2}{s_W^2}\frac{f_{M_2}}{f_{M_0}}\mathcal{M}^{f_{M_0}}(W^+\gamma\to W^+\gamma),\\
&\mathcal{M}^{f_{M_3}}(W^+\gamma\to W^+\gamma)=\frac{2c_W^2}{s_W^2}\frac{f_{M_3}}{f_{M_1}}\mathcal{M}^{f_{M_1}}(W^+\gamma\to W^+\gamma),\\
&\mathcal{M}^{f_{M_5}}(W^+\gamma\to W^+\gamma)=-\frac{2c_W}{s_W}\frac{f_{M_5}}{f_{M_1}}\mathcal{M}^{f_{M_1}}(W^+\gamma\to W^+\gamma),\\
&\mathcal{M}^{f_{M_7}}(W^+\gamma\to W^+\gamma)=-\frac{1}{2}\frac{f_{M_7}}{f_{M_1}}\mathcal{M}^{f_{M_1}}(W^+\gamma\to W^+\gamma),\\
&\mathcal{M}^{f_{T_5}}(W^+\gamma\to W^+\gamma)=\frac{c_W^2}{s_W^2}\frac{f_{T_5}}{f_{T_0}}\mathcal{M}^{f_{T_0}}(W^+\gamma\to W^+\gamma),\\
&\mathcal{M}^{f_{T_6}}(W^+\gamma\to W^+\gamma)=\frac{c_W^2}{s_W^2}\frac{f_{T_6}}{f_{T_1}}\mathcal{M}^{f_{T_1}}(W^+\gamma\to W^+\gamma),\\
&\mathcal{M}^{f_{T_7}}(W^+\gamma\to W^+\gamma)=\frac{c_W^2}{s_W^2}\frac{f_{T_7}}{f_{T_2}}\mathcal{M}^{f_{T_2}}(W^+\gamma\to W^+\gamma).\\
\end{split}
\label{eq.3.2}
\end{equation}
Therefore it is only necessary to calculate the partial-wave expansions of amplitudes for $O_{M_{0,1}}$ and $O_{T_{0,1,2}}$ operators.
The amplitudes grow with the c.m. energy $\sqrt{\hat{s}}$, keeping only the leading terms, the results are shown in Table~\ref{tab.awaw}.
There are also leading terms which can be obtained with the relation $\mathcal{M}_{\lambda _1, \lambda _2, \lambda _3, \lambda _4}(\theta ) = (-1)^{\lambda _1-\lambda _2-\lambda _3+\lambda _4}\mathcal{M}_{-\lambda _1, -\lambda _2, -\lambda _3, -\lambda _4}(\theta )$, therefore they are not presented.

\begin{table}
\caption{\label{tab.awaw}The partial-wave expansions of the $ W\gamma\to W\gamma$ amplitudes with one of dimension-8 operators $O_{M_{0,1}}$ and $O_{T_{0,1,2}}$ at the leading order.
The amplitudes set the strongest bounds are marked by a `*'. $\theta$ and $\varphi$ are zenith and azimuth angles of $\gamma$  in the final state.
}
\centering
\begin{tabular}{c|c|c}
\hline
amplitudes & leading order & expansions \\
\hline
$\mathcal{M}(\gamma _+W_0^+\to\gamma _-W_0^+)$ & $-\frac{f_{M_0}}{\Lambda^4}\frac{e^2  e^{i \varphi} v^2 \sin ^4\left(\frac{\theta}{2}\right)}{8 M_W^2}\hat{s}^2$ & $-\frac{f_{M_0}}{\Lambda^4}\frac{e^2  e^{2i \varphi} v^2}{8 M_W^2}\hat{s}^2\left(\frac{3}{4}d^1_{1,-1}-\frac{1}{4}d^2_{1,-1}\right)$ $\;^*$\\
& $\frac{f_{M_1}}{\Lambda^4}\frac{e^2  e^{i \varphi} v^2 \sin ^4\left(\frac{\theta}{2}\right)}{32 M_W^2}\hat{s}^2$ & $\frac{f_{M_1}}{\Lambda^4}\frac{e^2  e^{2i \varphi} v^2}{32 M_W^2}\hat{s}^2\left(\frac{3}{4}d^1_{1,-1}-\frac{1}{4}d^2_{1,-1}\right)$ \\
\hline
$\mathcal{M}(\gamma _+W_0^+\to\gamma _+W_0^+)$ & $\frac{f_{M_1}}{\Lambda^4}\frac{e^2 e^{i \varphi} v^2 (\cos (\theta)+1)}{32 M_W^2 }\hat{s}^2$ & $\frac{f_{M_1}}{\Lambda^4}\frac{e^2  v^2 }{16 M_W^2 }\hat{s}^2 d^1_{1,1}$ $\;^*$\\
\hline
$\mathcal{M}(\gamma _+W_+^+\to\gamma _-W_-^+)$ & $2\frac{f_{T_0}}{\Lambda^4} s_W^2 \sin ^4\left(\frac{\theta}{2}\right)\hat{s}^2$ & $2\frac{f_{T_0}}{\Lambda^4} s_W^2\hat{s}^2\left(\frac{1}{3}d^0_{0,0}-\frac{1}{2}d^1_{0,0}+\frac{1}{6}d^2_{0,0}\right)$\\
& $\frac{1}{2}\frac{f_{T_1}}{\Lambda^4}s_W^2\left(\sin ^4\left(\frac{\theta}{2}\right)+\left(\frac{\cos (\theta)+3}{2}\right)^2\right)\hat{s}^2$ & $\frac{1}{2}\frac{f_{T_1}}{\Lambda^4}s_W^2\hat{s}^2\left(-2d^0_{0,0}-2d^1_{0,0}\right)$ $\;^*$\\
& $\frac{1}{2}\frac{f_{T_2}}{\Lambda^4}s_W^2\sin ^4\left(\frac{\theta}{2}\right)\hat{s}^2$ & $\frac{1}{2}\frac{f_{T_2}}{\Lambda^4} s_W^2\hat{s}^2\left(\frac{1}{3}d^0_{0,0}-\frac{1}{2}d^1_{0,0}+\frac{1}{6}d^2_{0,0}\right)$\\
\hline
$\mathcal{M}(\gamma _+W_-^+\to\gamma _-W_+^+)$ & $2\frac{f_{T_0}}{\Lambda^4} e^{2 i \varphi} s_W^2 \sin ^4\left(\frac{\theta}{2}\right)\hat{s}^2$ & $2\frac{f_{T_0}}{\Lambda^4} e^{4 i \varphi} s_W^2 \hat{s}^2d^2_{2,-2}$ $\;^*$\\
& $\frac{1}{2}\frac{f_{T_2}}{\Lambda^4}e^{2 i \varphi} s_W^2 \sin ^4\left(\frac{\theta}{2}\right)\hat{s}^2$ & $\frac{1}{2}\frac{f_{T_2}}{\Lambda^4}e^{4 i \varphi} s_W^2 \hat{s}^2d^2_{2,-2}$\\
\hline
$\mathcal{M}(\gamma _-W_-^+\to\gamma _-W_-^+)$ & $\frac{f_{T_1}}{\Lambda^4}s_W^2\hat{s}^2$ & $\frac{f_{T_1}}{\Lambda^4}s_W^2\hat{s}^2d^0_{0,0}$ $\;^*$\\
& $\frac{1}{2}\frac{f_{T_2}}{\Lambda^4}s_W^2\hat{s}^2$ & $\frac{1}{2}\frac{f_{T_2}}{\Lambda^4}s_W^2\hat{s}^2d^0_{0,0}$ $\;^*$\\
\hline
$\mathcal{M}(\gamma _+W_-^+\to\gamma _+W_-^+)$ & $\frac{f_{T_1}}{\Lambda^4}e^{2i\varphi}s_W^2\cos ^4\left(\frac{\theta}{2}\right)\hat{s}^2$ & $\frac{f_{T_1}}{\Lambda^4}s_W^2\hat{s}^2d^2_{2,2}$\\
& $\frac{1}{2}\frac{f_{T_2}}{\Lambda^4}e^{2i\varphi}s_W^2\cos ^4\left(\frac{\theta}{2}\right)\hat{s}^2$ & $\frac{1}{2}\frac{f_{T_2}}{\Lambda^4}s_W^2\hat{s}^2d^2_{2,2}$\\
\hline
\end{tabular}
\end{table}

In Table~\ref{tab.awaw}, the channels with largest $|T_J|$ are marked with stars. From Table~\ref{tab.awaw} and Eq.~(\ref{eq.3.2}), we find the strongest bounds as
\begin{equation}
\begin{split}
\begin{array}{llll}
\left|\frac{f_{M_0}}{\Lambda^4}\right|\leq \frac{512 \pi M_W^2}{\hat{s}^2 e^2 v^2},\;
&\left|\frac{f_{M_1}}{\Lambda^4}\right|\leq \frac{768 \pi M_W^2}{e^2v^2\hat{s}^2},\;
&\left|\frac{f_{M_2}}{\Lambda^4}\right|\leq \frac{s_W^2 256 \pi M_W^2}{c_W^2e^2 v^2 \hat{s}^2},\;
&\left|\frac{f_{M_3}}{\Lambda^4}\right|\leq \frac{384 s_W^2 \pi M_W^2}{e^2v^2c_W^2\hat{s}^2},\\
\left|\frac{f_{M_4}}{\Lambda^4}\right|\leq \frac{s_W 512 \pi M_W^2}{c_We^2 v^2 \hat{s}^2},\;
&\left|\frac{f_{M_5}}{\Lambda^4}\right|\leq \frac{384s_W \pi M_W^2}{e^2v^2c_W\hat{s}^2},\;
&\left|\frac{f_{M_7}}{\Lambda^4}\right|\leq \frac{1536 \pi M_W^2}{e^2v^2\hat{s}^2},&\\
\left|\frac{f_{T_0}}{\Lambda^4}\right| \leq \frac{40\pi}{s_W^2 \hat{s}^2},\;
&\left|\frac{f_{T_1}}{\Lambda^4}\right| \leq \frac{32\pi}{s_W^2 \hat{s}^2},\;
&\left|\frac{f_{T_2}}{\Lambda^4}\right| \leq \frac{64\pi}{s_W^2\hat{s}^2},&\\
\left|\frac{f_{T_5}}{\Lambda^4}\right| \leq \frac{40\pi}{c_W^2 \hat{s}^2},\;
&\left|\frac{f_{T_6}}{\Lambda^4}\right| \leq \frac{32\pi}{c_W^2 \hat{s}^2},\;
&\left|\frac{f_{T_7}}{\Lambda^4}\right| \leq \frac{64\pi}{c_W^2\hat{s}^2}.&\\
\end{array}
\end{split}
\label{eq.3.3}
\end{equation}

\subsection{\label{level3.2}Partial-wave expansions of the \texorpdfstring{$WZ\to W\gamma$}{WZ to WA} amplitudes}
\begin{table}[!htbp]
\caption{\label{tab.zwaw} Same as Table~\ref{tab.awaw} but for $WZ\to W\gamma$.}
\centering
\begin{tabular}{c|c|c}
\hline
amplitudes  & leading order & expansions \\
\hline
$\mathcal{M}(Z_+W_0^+\to\gamma _-W_0^+)$ & $-\frac{f_{M_0}}{\Lambda^4}\frac{c_W e^2 e^{i \varphi} v^2 \sin ^4\left(\frac{\theta}{2}\right) }{8 M_W^2 s_W}\hat{s}^2$ & $-\frac{f_{M_0}}{\Lambda^4}\frac{c_W e^2 e^{2i \varphi} v^2  }{8 M_W^2 s_W}\hat{s}^2\left(\frac{3}{4}d^1_{1,-1}-\frac{1}{4}d^2_{1,-1}\right)$ $\;^*$\\
& $\frac{f_{M_1}}{\Lambda^4}\frac{c_W e^2 e^{i \varphi} v^2 \sin ^4\left(\frac{\theta}{2}\right) }{32 M_W^2 s_W}\hat{s}^2$  & $\frac{f_{M_1}}{\Lambda^4}\frac{c_W e^2 e^{2i \varphi} v^2 }{32 M_W^2 s_W}\hat{s}^2\left(\frac{3}{4}d^1_{1,-1}-\frac{1}{4}d^2_{1,-1}\right)$\\
& $\frac{f_{M_4}}{\Lambda^4}\frac{e^2 e^{i \varphi} v^2 \left(s_W^2-c_W^2\right) \sin ^4\left(\frac{\theta}{2}\right) }{16 M_W^2 c_W^2}\hat{s}^2$  & $\frac{f_{M_4}}{\Lambda^4}\frac{e^2 e^{2i \varphi} v^2 \left(s_W^2-c_W^2\right) }{16 M_W^2 c_W^2}\hat{s}^2\left(\frac{3}{4}d^1_{1,-1}-\frac{1}{4}d^2_{1,-1}\right)$\\
& $\frac{f_{M_5}}{\Lambda^4}\frac{e^2 e^{i \varphi} v^2 \left(s_W^2-c_W^2\right) \sin ^4\left(\frac{\theta}{2}\right) }{32 M_W^2 s_W^2}\hat{s}^2$  & $\frac{f_{M_5}}{\Lambda^4}\frac{e^2 e^{2i \varphi} v^2 \left(s_W^2-c_W^2\right)  }{32 M_W^2 s_W^2}\hat{s}^2\left(\frac{3}{4}d^1_{1,-1}-\frac{1}{4}d^2_{1,-1}\right)$ \\
& $-\frac{f_{M_7}}{\Lambda^4}\frac{c_W e^2 e^{i \varphi} v^2 \sin ^4\left(\frac{\theta}{2}\right) }{64 M_W^2s_W}\hat{s}^2$  & $-\frac{f_{M_7}}{\Lambda^4}\frac{c_W e^2 e^{2i \varphi} v^2 }{64 M_W^2s_W}\hat{s}^2\left(\frac{3}{4}d^1_{1,-1}-\frac{1}{4}d^2_{1,-1}\right)$\\
\hline
$\mathcal{M}(Z_+W_0^+\to\gamma _+W_0^+)$ & $\frac{f_{M_1}}{\Lambda^4}\frac{c_W e^2 e^{i \varphi} v^2 \cos ^2\left(\frac{\theta}{2}\right)}{16 M_W^2s_W}\hat{s}^2$ & $\frac{f_{M_1}}{\Lambda^4}\frac{c_W e^2  v^2 }{16 M_W^2s_W}\hat{s}^2d^1_{1,1}$ $\;^*$\\
& $\frac{f_{M_5}}{\Lambda^4}\frac{e^2 e^{i \varphi} v^2 \left(s_W^2-c_W^2\right) \cos ^2\left(\frac{\theta}{2}\right)}{16 M_W^2 s_W^2}\hat{s}^2$ & $\frac{f_{M_5}}{\Lambda^4}\frac{e^2  v^2 \left(s_W^2-c_W^2\right) }{16 M_W^2 s_W^2}\hat{s}^2d^1_{1,1}$\\
& $-\frac{f_{M_7}}{\Lambda^4}\frac{c_W e^2 e^{i \varphi} v^2 \cos ^2\left(\frac{\theta}{2}\right)}{32M_W^2s_W}\hat{s}^2$ & $-\frac{f_{M_7}}{\Lambda^4}\frac{c_W e^2  v^2 }{32M_W^2s_W}\hat{s}^2d^1_{1,1}$  $\;^*$\\
\hline
$\mathcal{M}(Z_0W_+^+\to\gamma _-W_0^+)$ & $\frac{f_{M_4}}{\Lambda^4}\frac{e^2 e^{-i \varphi} v^2 \cos ^4\left(\frac{\theta}{2}\right) }{16 M_WM_Zs_W^2}\hat{s}^2$ & $\frac{f_{M_4}}{\Lambda^4}\frac{e^2v^2}{64 M_WM_Zs_W^2}\hat{s}^2(3d^1_{-1,-1}+d^2_{-1,-1})$ \\
& $\frac{f_{M_5}}{\Lambda^4}\frac{e^2 e^{-i \varphi} v^2 \cos ^4\left(\frac{\theta}{2}\right)}{32 M_WM_Zs_W^2}\hat{s}^2$ & $\frac{f_{M_5}}{\Lambda^4}\frac{e^2v^2 }{128 M_WM_Zs_W^2}\hat{s}^2(3d^1_{-1,-1}+d^2_{-1,-1})$ \\
& $\frac{f_{M_7}}{\Lambda^4}\frac{e^2 e^{-i \varphi} v^2 \cos ^2\left(\frac{\theta}{2}\right) (\cos (\theta)-3)}{128 c_W s_W M_WM_Z}\hat{s}^2$ & $\frac{f_{M_7}}{\Lambda^4}\frac{e^2 v^2 }{256 c_W s_W M_WM_Z}\hat{s}^2(-5d^1_{-1,-1}+d^2_{-1,-1})$ \\
\hline
$\mathcal{M}(Z_0W_0^+\to\gamma _+W_+^+)$ & $\frac{f_{M_4}}{\Lambda^4}\frac{e^2 v^2 }{16 M_WM_Z s_W^2}\hat{s}^2$ & $\frac{f_{M_4}}{\Lambda^4}\frac{e^2 v^2}{16 M_WM_Z s_W^2}\hat{s}^2d^0_{0,0}$ $\;^*$ \\
& $\frac{f_{M_5}}{\Lambda^4}\frac{e^2 v^2 }{32 M_WM_Z s_W^2}\hat{s}^2$ & $\frac{f_{M_5}}{\Lambda^4}\frac{e^2 v^2}{32 M_WM_Z s_W^2}\hat{s}^2d^0_{0,0}$  $\;^*$\\
& $-\frac{f_{M_7}}{\Lambda^4}\frac{e^2 v^2 \cos (\theta)}{64 c_Ws_WM_WM_W}\hat{s}^2$ & $-\frac{f_{M_7}}{\Lambda^4}\frac{e^2 v^2}{64 c_Ws_WM_WM_W}\hat{s}^2d^1_{0,0}$ \\
\hline
$\mathcal{M}(Z_0W_0^+\to\gamma _+W_-^+)$ & $-\frac{f_{M_5}}{\Lambda^4}\frac{e^2 v^2 \sin ^2(\theta)}{64 M_WM_Z s_W^2}\hat{s}^2$ & $-\frac{f_{M_5}}{\Lambda^4}\frac{e^2 v^2 e^{-2i\varphi}}{32 M_WM_Z s_W^2}\hat{s}^2\sqrt{\frac{2}{3}}d^2_{0,2}$ \\
\hline
$\mathcal{M}(Z_+W_+^+\to\gamma _-W_-^+)$ & $2\frac{f_{T_0}}{\Lambda^4}c_Ws_W \sin ^4\left(\frac{\theta}{2}\right)\hat{s}^2$ & $2\frac{f_{T_0}}{\Lambda^4}c_Ws_W \hat{s}^2\left(\frac{1}{3}d^0_{0,0}-\frac{1}{2}d^1_{0,0}+\frac{1}{6}d^2_{0,0}\right)$ \\
& $\frac{f_{T_1}}{\Lambda^4}c_Ws_W \frac{4 \cos (\theta)+\cos (2 \theta)+11}{8}\hat{s}^2$ & $\frac{1}{2}\frac{f_{T_1}}{\Lambda^4}c_Ws_W \hat{s}^2\left(\frac{8}{3}d^0_{0,0}+d^1_{0,0}+\frac{1}{3}d^2_{0,0}\right)$ $\;^*$\\
& $\frac{f_{T_2}}{\Lambda^4}c_Ws_W \frac{\cos (2 \theta)-4 \cos (\theta)+3}{16}\hat{s}^2$ & $\frac{1}{4}\frac{f_{T_2}}{\Lambda^4}c_Ws_W \hat{s}^2\left(\frac{2}{3}d^0_{0,0}-d^1_{0,0}+\frac{1}{3}d^2_{0,0}\right)$\\
\hline
$\mathcal{M}(Z_+W_-^+\to\gamma _-W_+^+)$ & $2\frac{f_{T_0}}{\Lambda^4}c_Ws_We^{2i\varphi}\sin ^4 \left(\frac{\theta}{2}\right)\hat{s}^2$& $2\frac{f_{T_0}}{\Lambda^4}c_Ws_We^{4i\varphi}\hat{s}^2d^2_{2,-2}$ $\;^*$\\
& $\frac{1}{2}\frac{f_{T_2}}{\Lambda^4}c_Ws_We^{2i\varphi}\sin ^4 \left(\frac{\theta}{2}\right)\hat{s}^2$ & $\frac{1}{2}\frac{f_{T_2}}{\Lambda^4}c_Ws_We^{4i\varphi}s^2d^2_{2,-2}$ \\
\hline
$\mathcal{M}(Z_+W_+^+\to\gamma _+W_+^+)$ & $\frac{f_{T_1}}{\Lambda^4}c_Ws_W\hat{s}^2$ & $\frac{f_{T_1}}{\Lambda^4}c_Ws_W\hat{s}^2d^0_{0,0}$\\
& $\frac{1}{2}\frac{f_{T_2}}{\Lambda^4}c_Ws_W\hat{s}^2$ & $\frac{1}{2}\frac{f_{T_2}}{\Lambda^4}c_Ws_W\hat{s}^2d^0_{0,0}$ $\;^*$\\
\hline
$\mathcal{M}(Z_+W_-^+\to\gamma _+W_-^+)$ & $\frac{f_{T_1}}{\Lambda^4}c_Ws_We^{2i\varphi}\cos ^4 \left(\frac{\theta}{2}\right)\hat{s}^2$ & $\frac{f_{T_1}}{\Lambda^4}c_Ws_W\hat{s}^2d^2_{2,2}$\\
& $\frac{1}{2}\frac{f_{T_2}}{\Lambda^4}c_Ws_We^{2i\varphi}\cos ^4 \left(\frac{\theta}{2}\right)\hat{s}^2$ & $\frac{1}{2}\frac{f_{T_2}}{\Lambda^4}c_Ws_W\hat{s}^2d^2_{2,2}$ \\
\hline
\end{tabular}
\end{table}

For $WZ\to W\gamma$, similarly,
\begin{equation}
\begin{split}
&\mathcal{M}^{f_{M_2}}(W^+Z\to W^+\gamma)=-2\frac{f_{M_2}}{f_{M_0}}\mathcal{M}^{f_{M_0}}(W^+Z\to W^+\gamma),\\
&\mathcal{M}^{f_{M_3}}(W^+Z\to W^+\gamma)=-2\frac{f_{M_3}}{f_{M_1}}\mathcal{M}^{f_{M_1}}(W^+Z\to W^+\gamma),\\
&\mathcal{M}^{f_{T_5}}(W^+Z\to W^+\gamma)=-\frac{f_{T_5}}{f_{T_0}}\mathcal{M}^{f_{T_0}}(W^+Z\to W^+\gamma),\\
&\mathcal{M}^{f_{T_6}}(W^+Z\to W^+\gamma)=-\frac{f_{T_6}}{f_{T_1}}\mathcal{M}^{f_{T_1}}(W^+Z\to W^+\gamma),\\
&\mathcal{M}^{f_{T_7}}(W^+Z\to W^+\gamma)=-\frac{f_{T_7}}{f_{T_2}}\mathcal{M}^{f_{T_2}}(W^+Z\to W^+\gamma).\\
\end{split}
\label{eq.3.4}
\end{equation}

Then, the partial-wave expansions for the amplitudes of $O_{M_{0,1,4,5,7}}$ and $O_{T_{0,1,2}}$ are shown in Table~\ref{tab.zwaw}.
The strongest bounds can be obtained via Table~\ref{tab.zwaw} and Eq.~(\ref{eq.3.4}),
\begin{equation}
\begin{split}
&\left|\frac{f_{M_0}}{\Lambda^4}\right|\leq \frac{512\pi M_W^2 s_W}{c_W e^2 v^2 \hat{s}^2},\;
 \left|\frac{f_{M_1}}{\Lambda^4}\right|\leq \frac{768 \pi M_W^2 s_W}{c_W e^2 v^2 \hat{s}^2},\;
 \left|\frac{f_{M_2}}{\Lambda^4}\right|\leq \frac{256\pi M_W^2 s_W}{c_W e^2 v^2 \hat{s}^2},\;
 \left|\frac{f_{M_3}}{\Lambda^4}\right|\leq \frac{384 \pi M_W^2s_W}{c_W e^2 v^2 \hat{s}^2},\\
&\left|\frac{f_{M_4}}{\Lambda^4}\right|\leq \frac{512 \pi M_WM_Z s_W^2}{e^2v^2\hat{s}^2},\;
 \left|\frac{f_{M_5}}{\Lambda^4}\right|\leq \frac{1024\pi M_WM_Zs_W^2}{e^2v^2\hat{s}^2},\;
 \left|\frac{f_{M_7}}{\Lambda^4}\right|\leq \frac{1536\pi M_W^2 s_W}{e^2v^2 c_W\hat{s}^2},\\
&\left|\frac{f_{T_0}}{\Lambda^4}\right|\leq \frac{40\pi}{c_Ws_W\hat{s}^2},\;
 \left|\frac{f_{T_1}}{\Lambda^4}\right|\leq \frac{24\pi}{c_Ws_W\hat{s}^2},\;
 \left|\frac{f_{T_2}}{\Lambda^4}\right|\leq \frac{64\pi}{c_Ws_W\hat{s}^2},\\
&\left|\frac{f_{T_5}}{\Lambda^4}\right|\leq \frac{40\pi}{c_Ws_W\hat{s}^2},\;
 \left|\frac{f_{T_6}}{\Lambda^4}\right|\leq \frac{24\pi}{c_Ws_W\hat{s}^2},\;
 \left|\frac{f_{T_7}}{\Lambda^4}\right|\leq \frac{64\pi}{c_Ws_W\hat{s}^2},\\
\end{split}
\label{eq.3.5}
\end{equation}

\subsection{\label{level3.3}Partial-wave unitarity bounds}

For $W\gamma jj$ production, the process $W\gamma\to W\gamma$ cannot be distinguished from the process $WZ\to W\gamma$. Therefore, we set the unitarity bounds by requiring all events to satisfy the strongest bounds.
From Eqs.~(\ref{eq.3.3}) and (\ref{eq.3.5}), the strongest bounds are given by
\begin{equation}
\begin{split}
&\left|\frac{f_{M_0}}{\Lambda^4}\right|\leq \frac{512 \pi M_W^2s_W}{c_We^2v^2\hat{s}^2},\;
 \left|\frac{f_{M_1}}{\Lambda^4}\right|\leq \frac{768 \pi M_W^2s_W}{c_We^2v^2\hat{s}^2},\;
 \left|\frac{f_{M_2}}{\Lambda^4}\right|\leq \frac{s_W^2 256 \pi M_W^2}{c_W^2e^2 v^2 \hat{s}^2},\;
 \left|\frac{f_{M_3}}{\Lambda^4}\right|\leq \frac{384 \pi s_W^2 M_W^2}{c_W^2e^2v^2\hat{s}^2},\\
&\left|\frac{f_{M_4}}{\Lambda^4}\right|\leq \frac{512 \pi M_WM_Zs_W^2}{e^2 v^2 \hat{s}^2},\;
 \left|\frac{f_{M_5}}{\Lambda^4}\right|\leq \frac{384\pi M_WM_Zs_W}{c_We^2v^2\hat{s}^2},\;
 \left|\frac{f_{M_7}}{\Lambda^4}\right|\leq \frac{1536 s_W\pi M_W^2}{e^2v^2c_W\hat{s}^2},\\
&\left|\frac{f_{T_0}}{\Lambda^4}\right| \leq \frac{40\pi}{s_Wc_W \hat{s}^2},\;
 \left|\frac{f_{T_1}}{\Lambda^4}\right| \leq \frac{24\pi}{s_Wc_W \hat{s}^2},\;
 \left|\frac{f_{T_2}}{\Lambda^4}\right| \leq \frac{64\pi}{s_Wc_W \hat{s}^2},\\
&\left|\frac{f_{T_5}}{\Lambda^4}\right| \leq \frac{40\pi}{c_W^2 \hat{s}^2},\;
 \left|\frac{f_{T_6}}{\Lambda^4}\right| \leq \frac{32\pi}{c_W^2 \hat{s}^2},\;
 \left|\frac{f_{T_7}}{\Lambda^4}\right| \leq \frac{64\pi}{c_W^2\hat{s}^2}.\\
\end{split}
\label{eq.3.6}
\end{equation}

The unitarity bounds indicate that the events with large enough $\sqrt{\hat{s}}$ could not be described by the SMEFT correctly.
The violation of unitarity can be avoided by unitarization methods such as K-matrix unitarization~\cite{kmatrix} or by putting form factors into the coefficients~\cite{VBSReview,VBFCut}, as well as via dispersion relations~\cite{dispersion}.
It is pointed out that the constraints on the effective couplings dependent on the method used, one should not relay on just one-method~\cite{UnitarizationWZ}.
On the other hand, in experiments, the constraints on the coefficients are obtained with the EFT without unitarization. To compare with the experimental data, we present our results without unitization in this paper.

In VBS processes, the initial states are protons, therefore $\sqrt{\hat{s}}$ is a distribution related to the parton distribution function of proton, one cannot set the constraints on the coefficients by $\hat{s}$.
In this work, we discarded the events with large $\hat{s}$ to ensure the events generated by the SMEFT are in the valid region.
In other words, we compare the signals of aQGCs with the backgrounds under a certain energy scale cut similar as the matching procedure in Refs.~\cite{shatcut,ecut}.

\section{\label{level4}The signals of aQGCs and the backgrounds}

\begin{figure}
\begin{center}
\includegraphics[width=0.8\textwidth]{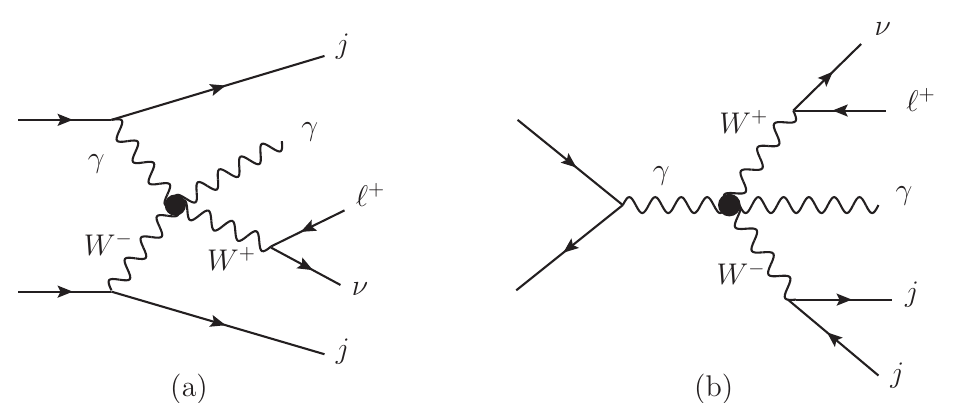}
\caption{The typical aQGC diagrams contributing to $\ell^+\nu\gamma jj$ final states. Similar as in the SM, there are also VBS contributions as depicted in (a) and non-VBS contributions as shown in (b).
}
\label{fig:typicalSignal}
\end{center}
\end{figure}

\begin{figure}
\begin{center}
\includegraphics[width=0.8\textwidth]{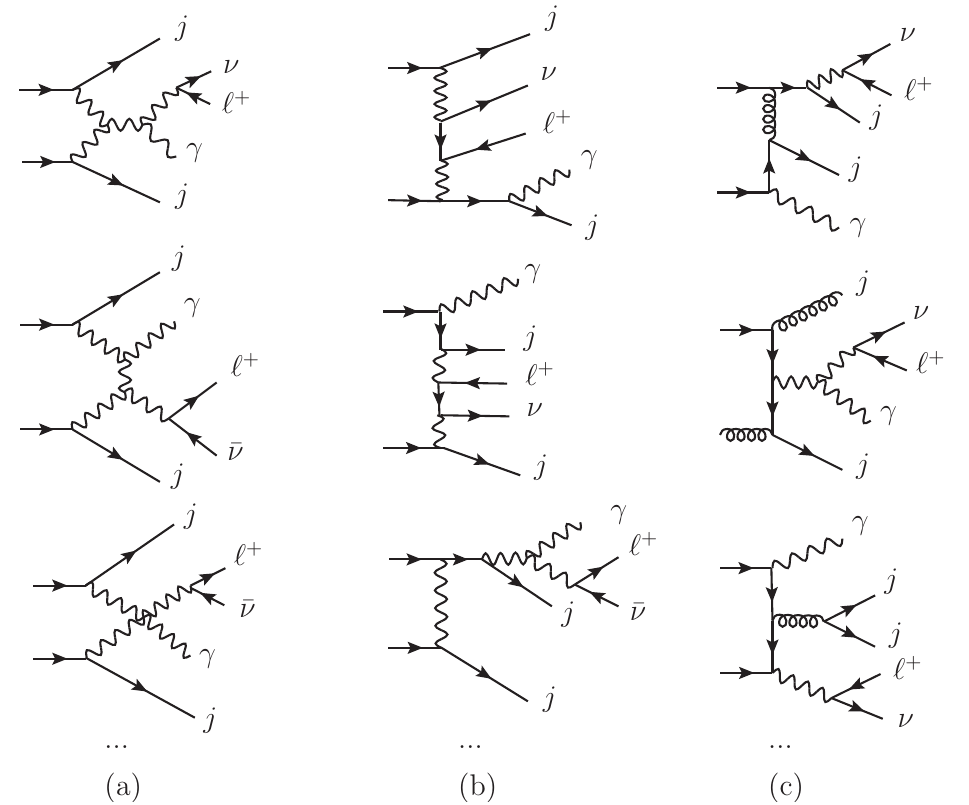}
\caption{
The typical Feynman diagrams of the SM backgrounds including the EW-VBS (a), EW-non-VBS (b) and QCD diagrams (c).
}
\label{fig:typicalBackground}
\end{center}
\end{figure}

The dominant signal is leptonic decay of $W\gamma jj$ production induced by the dimension-8 operators, and we consider one operator at a time. The Feynman diagrams are shown in Fig.~\ref{fig:typicalSignal}.~(a).
The triboson diagrams such as Fig.~\ref{fig:typicalSignal}.~(b) also contribute to the signal.
The typical Feynman diagrams of the SM backgrounds can be found in Fig.~\ref{fig:typicalBackground}, which are often categorized as the EW VBS, EW non-VBS and QCD contributions.
The triboson contribution from each $O_{M_i}$ ($O_{T_i}$) operator is two (three) orders of magnitude smaller than the dominant signal even with the interferences considered. So we concentrate on the dominant signal.


The numerical results are obtained by Monte-Carlo~(MC) simulation using \verb"MadGraph5_aMC@NLO"~(MG5) toolkit~\cite{madgraph}.
The parton distribution function is NNPDF2.3~\cite{NNPDF}. The renormalization scale $\mu_r$ and factorization scale $\mu_f$ are
chosen to be dynamical which are set event-by-event as $\left(\prod _i^n \left(M_i^2+ {\vec p}_T^{(i)}\right)\right)^{\frac{1}{n}}$, with $i$ running over all heavy particles.
The basic cuts are applied with the default settings of MG5 which require ${\vec p}_T^{\gamma,\ell}>10$ GeV, $|\eta _{\gamma,\ell}|>2.5$ and ${\vec p}_T^{j}>20$ GeV, $|\eta _j| < 5$.
The events are then showered by \verb"PYTHIA8" \cite{pythia} and a fast detector simulation are performed by \verb"Delphes" \cite{delphes} with the CMS detector card.
Jets are clustered by using the anti-$k_T$ algorithm with a cone radius $R=0.5$ and $\vec{p}_{T,min}=20$ GeV.
The photon isolation uses parameter $I_{min}$ defined as \cite{delphes}
\begin{equation}
I_{min}^{\gamma}=\frac{\sum _{i\neq \gamma}^{\Delta R<\Delta R_{max}, {\vec p}_T^{i}>{\vec p}_{T,min}} {\vec p}^{i}_T}{{\vec p}_T^{\gamma}},
\end{equation}
where $\Delta R_{max}=0.5$, $\vec{p}_{T,min}=0.5$ GeV. $I_{min}^{\gamma}>0.12$ is required.
We generate the dominant signal events with the largest coefficients in Table~\ref{tab.1}. 
The analyses of the signal and the background are completed by \verb"MLAnalysis"~\cite{Guo:2023nfu}. 
After fast detector simulation, the final states are not exactly $\ell ^+ \nu \gamma jj$.
To ensure a high quality track of the signal candidate, a minimum number of composition is required.
We denote the number of jets, photons and charged leptons as $N_j$, $N_{\gamma}$ and $N_{\ell^+}$, respectively.
Events are selected by requiring $N_j\geq 2$, $N_{\gamma}\geq 1$ and $N_{\ell^+}=1$.
We analyse the energy scale, kinematic features and polarization features of the events after these particle number cuts.

Since the $O_{M_{0,1,7}}$ and $O_{T_{0,1,2}}$ operators are constrained tightly by $WWjj$, $WZjj$, and $ZZjj$ productions~\cite{newdatawvzv}, we concentrate on the $O_{M_{2,3,4,5}}$ and $O_{T_{5,6,7}}$ operators.

\subsection{\label{level4.0} Implementation of unitarity bounds}

\begin{figure}
\includegraphics[width=0.49\textwidth]{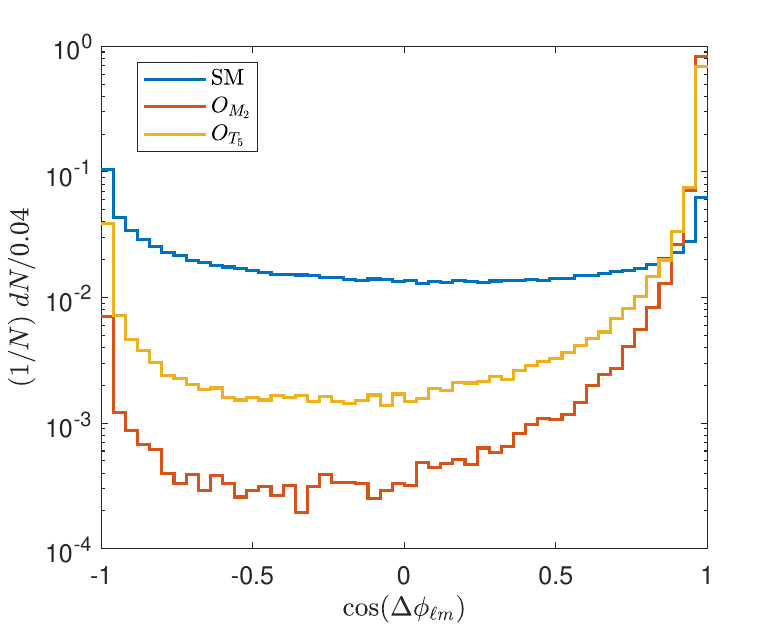}
\includegraphics[width=0.49\textwidth]{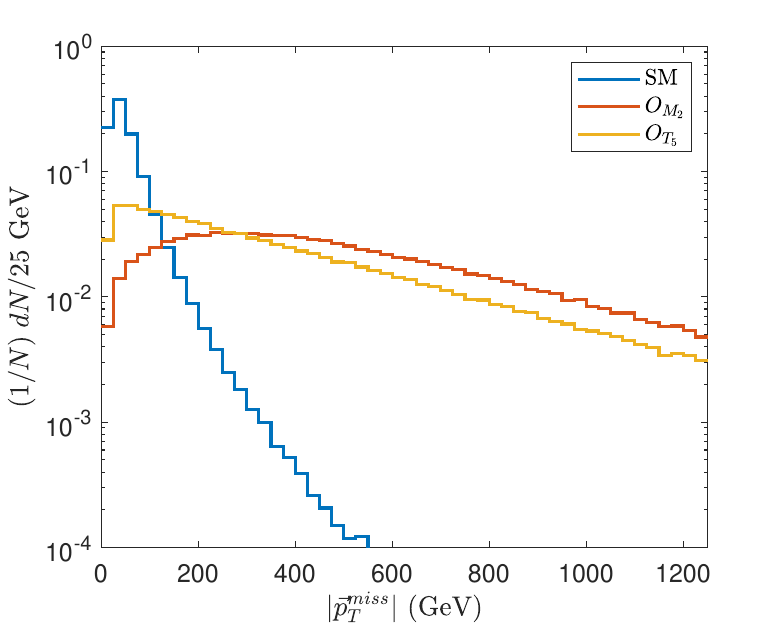}
\includegraphics[width=0.5\textwidth]{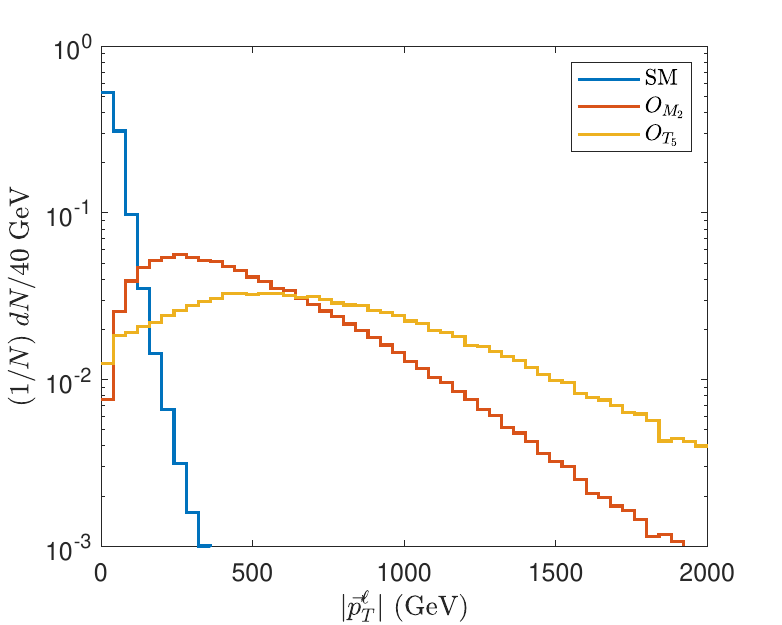}
\caption{The normalized distributions of $\cos (\Delta \phi _{\ell m})$, $|\vec{p}_T^{miss}|$ and $|\vec{p}_T^{\ell}|$ after the particle number cuts.}
\label{fig:philm}
\end{figure}

To make sure the events are generated by the EFT in a valid region, the unitarity bounds are applied as cuts on $\hat{s}$.
However, $\hat{s}$ is not an observable because of the invisible neutrino. Instead, we find an observable to evaluate $\hat{s}$ approximately.
We use the approximation that most of the $W$ bosons are on shell, such that $(p_{\ell}+p_{\nu})^2\approx M_W^2\ll \hat{s}$.
Compared with a large $\hat{s}$, the mass of the $W$ boson is negligible, thus $2p_{\ell} p_{\nu}\approx 0$, which indicates that the flight direction of the neutrino is close to the charged lepton.
We use an event selection strategy to select the events with small azimuth angle between the charged lepton and the missing momentum which is denoted as $\Delta \phi _{\ell m}$ to strengthen this approximation. The normalized distributions of $\cos (\Delta \phi _{\ell m})$ are shown in Fig.~\ref{fig:philm}.~(a). The distributions are similar for each class of operators (i.e. $O_{M_i}$ or $O_{T_i}$), but are different between $O_{M_i}$ and $O_{T_i}$.
Therefore we only present $O_{M_2}$ and $O_{T_5}$ as examples. We choose $\cos (\Delta \phi _{\ell m})>0.95$ to cut off the events with a small $\cos (\Delta \phi _{\ell m})$.

Using the approximation that neutrino and the charged lepton are nearly parallel to each other, and by also requiring $|{\vec p}^{\ell}_T|>0$ which is guaranteed due to the detector simulation, we introduce
\begin{equation}
\begin{split}
\tilde{s}&= \left(\sqrt{|\vec{p}_T^{miss}|^2+\left(\frac{|\vec{p}^{miss}_T|}{|\vec{p}_T^{\ell}|}p^{\ell}_z\right)^2}+E_{\ell}+E_{\gamma}\right)^2\\
&-\left(\left(1+\frac{|\vec{p}^{miss}_T|}{|\vec{p}_T^{\ell}|}\right)p_z^{\ell}+p_z^{\gamma}\right)^2-\left|\vec{p}_T^{\ell}+\vec{p}_T^{miss}+\vec{p}_T^{\gamma}\right|^2,\\
\end{split}
\label{eq.4.1}
\end{equation}
where $E_{\ell,\gamma}=\sqrt{(p_x^{\ell,\gamma})^2+(p_y^{\ell,\gamma})^2+(p_z^{\ell,\gamma})^2}$, $\vec{p}_T^{\ell}=(p_x^{\ell},p_y^{\ell},0)$, $p_{x,y,z}^{\ell,\gamma}$ are components of the momenta of lepton and photon in the c.m. frame of $pp$, $\vec{p}^{miss}_T$ is the missing momentum. $\tilde{s}$ reconstruct $\hat{s}$ when neutrino and the charged lepton are exactly collinear and when the missing momentum is exactly neutrino transverse momentum.
From the definition of $\tilde{s}$, one can see that, with a larger $|\vec{p}^{\ell}_T|$, the approximation is better. Meanwhile, the cross sections of the sub-processes $W^+\gamma\to W^+\gamma$ and $ZW^+\to W^+\gamma$ grow with $\sqrt{\hat{s}}$, therefore one can expect that the number of signal events grows with increasing $\hat{s}$, namely, one can expect an energetic $W^+$ boson, therefore the momentum of the charged lepton produced by the $W^+$ boson should also be large.
For the same reason, $|\vec{p}_T^{miss}|$ should also be large.
A small $|\vec{p}_T^{miss}|$  probably indicates a neutrino alone the $\vec{z}$ direction, the approximation $\tilde{s}$ can benefit from cutting off such events.
The normalized distributions of $|\vec{p}_T^{\ell}|$ and $|\vec{p}_T^{miss}|$ after particle number cuts are shown in Fig.~\ref{fig:philm}.~(b) and (c). We choose the events with $|\vec{p}_T^{\ell}|> 80\;{\rm GeV}$ and $|\vec{p}_T^{miss}|> 50\;{\rm GeV}$.

\begin{figure}
\subfloat[$O_{M_2}$]{\includegraphics[width=0.5\textwidth]{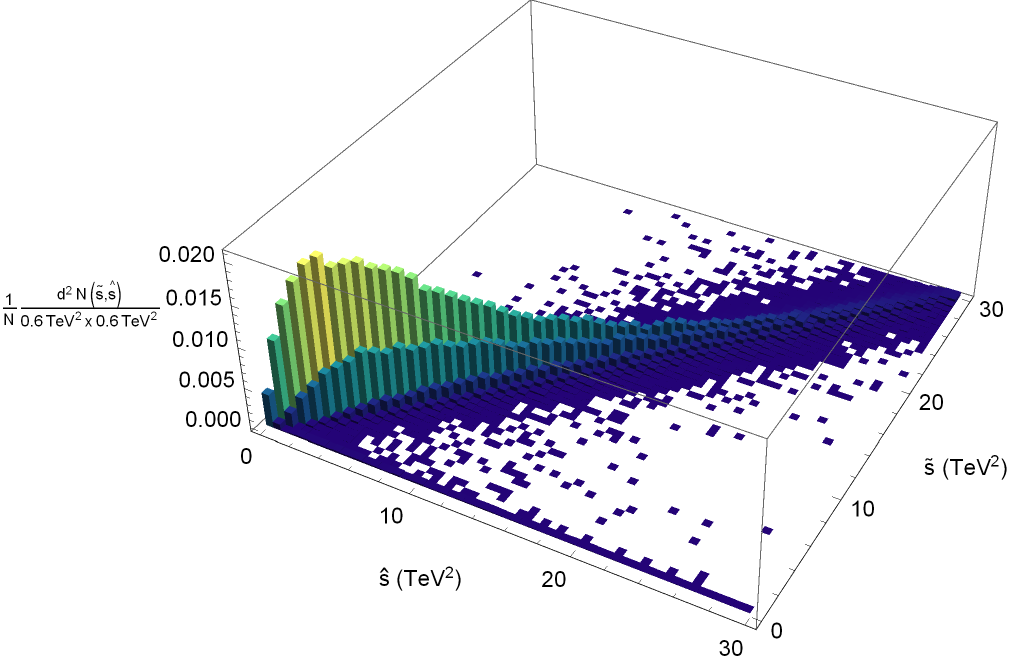}}\hfill
\subfloat[$O_{T_5}$]{\includegraphics[width=0.5\textwidth]{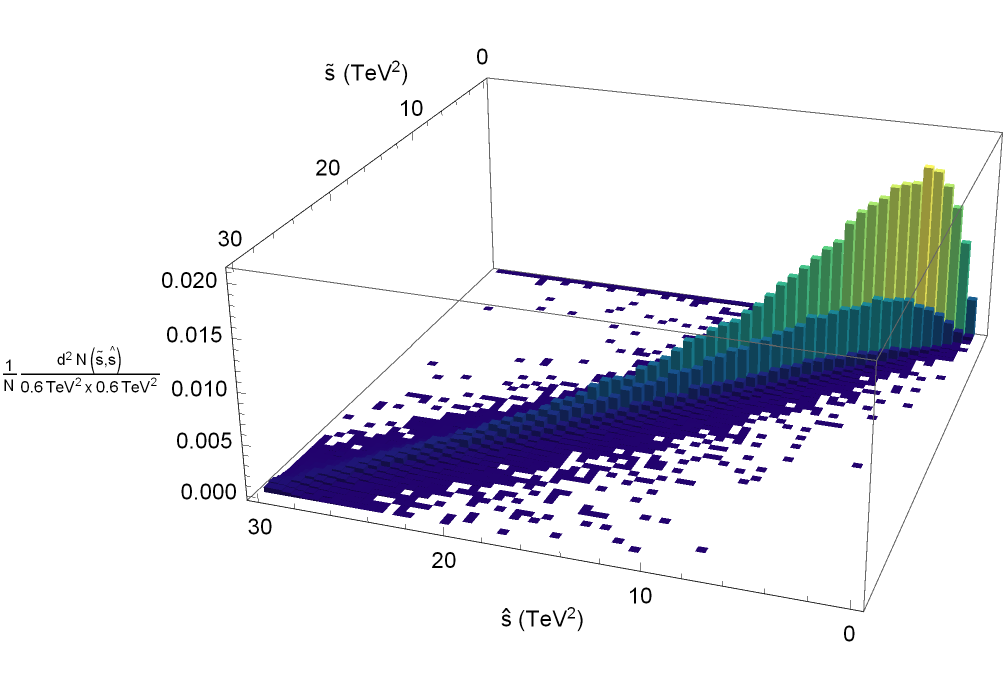}}\vfill
\subfloat[$\tilde{s}-\hat{s}$]{\includegraphics[width=0.6\textwidth]{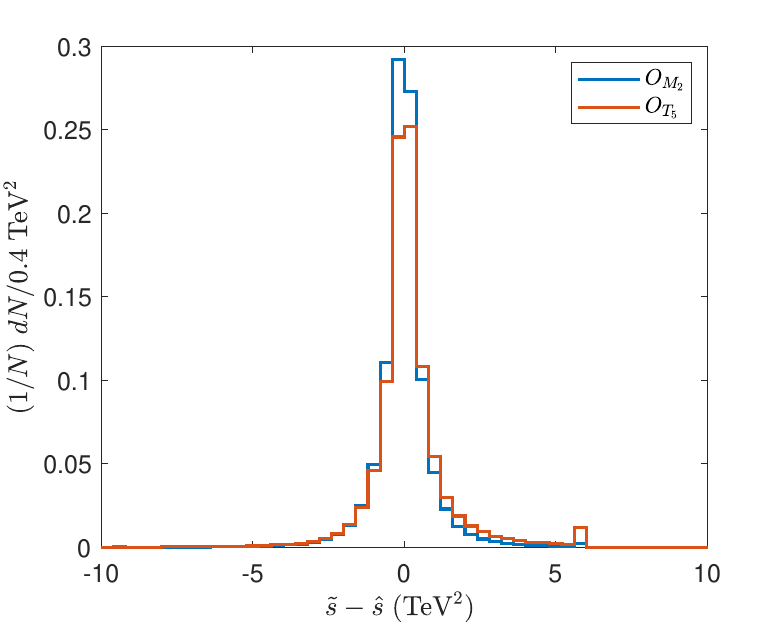}}\hfill
\caption{The correlation between $\hat{s}$ and $\tilde{s}$ for $O_{M_2}$ and $O_{T_5}$.}
\label{fig:shat}
\end{figure}

To verify the approximation accuracy, we calculate both $\hat{s}$ and $\tilde{s}$. Unlike real experiments, in simulation $\hat{s}$ can be obtained before detector simulation.
Both $\hat{s}$ and $\tilde{s}$ are calculated after the $\Delta \phi _{\ell m}$, $|\vec{p}_T^{\ell}|$ and $|\vec{p}_T^{miss}|$ cuts are applied.
Take $O_{M_2}$ and $O_{T_5}$ operators for example, as shown in Fig.~\ref{fig:shat}, $\tilde{s}$ can approximate $\hat{s}$ well.

The unitarity bounds are realized as energy cuts using $\tilde{s}$, denoted as $\tilde{s}_U$. From Eq.~(\ref{eq.3.6}), the $\tilde{s}_U$ cuts are
\begin{equation}
\begin{split}
&\tilde{s}(f_{M_2}) \leq \sqrt{\frac{s_W^2 256 \pi M_W^2 \Lambda^4}{c_W^2e^2 v^2 |f_{M_2}|}},\;\;
 \tilde{s}(f_{M_3}) \leq \sqrt{\frac{384 \pi s_W^2 M_W^2 \Lambda^4}{c_W^2e^2v^2 |f_{M_3}|}},\\
&\tilde{s}(f_{M_4}) \leq \sqrt{\frac{512 \pi M_WM_Zs_W^2 \Lambda^4}{e^2 v^2 |f_{M_4}|}},\;\;
 \tilde{s}(f_{M_5}) \leq \sqrt{\frac{384\pi M_WM_Zs_W \Lambda^4}{c_We^2v^2 |f_{M_5}|}},\\
&\tilde{s}(f_{T_5}) \leq \sqrt{\frac{40\pi \Lambda^4}{c_W^2 |f_{T_5}|}},\;\;
 \tilde{s}(f_{T_6}) \leq \sqrt{\frac{32\pi \Lambda^4}{c_W^2 |f_{T_6}|}},\;\;
 \tilde{s}(f_{T_7}) \leq \sqrt{\frac{64\pi \Lambda^4}{c_W^2 |f_{T_7}|}}.\\
\end{split}
\label{eq.4.2}
\end{equation}
The effects of the $\Delta \phi _{\ell m}$, $|\vec{p}_T^{\ell}|$, $|\vec{p}_T^{miss}|$ and $\tilde{s}_U$ cuts are shown in Table~\ref{tab.unitaritybound}. Theoretically, the unitarity bounds should not be applied to the SM backgrounds. However, in the aspect of experiment we cannot distinguish the aQGC signals from the SM backgrounds strictly, thus the $\tilde{s}_U$ cut can only be applied on all events. Therefore, we also apply the $\tilde{s}_U$ cuts on the SM backgrounds. We verify that the $\tilde{s}_U$ cuts have negligible effects on the SM backgrounds for all largest $f_{M_{2,3,4,5}}/\Lambda^4$ and $f_{T_{5,6,7}}/\Lambda^4$ we are using.

From Table~\ref{tab.unitaritybound}, we find that the unitarity bounds have significant suppressive effects on the signals, especially for $O_{M_i}$ operators, indicating the necessity of the unitarity bounds.

\begin{table}
\caption{\label{tab.unitaritybound}The cross sections of the SM backgrounds and signals for different operators after $N_{j,\gamma,\ell^+}$, $\Delta \phi _{\ell m}$, $|\vec{p}^{\ell}_T|$, $|\vec{p}_T^{miss}|$ and $\tilde{s}_U$ cuts.
The maximum $\tilde{s}$ used in the $\tilde{s}_U$ cuts are obtained by the upper bounds of $f_X/\Lambda ^4$ in Table~\ref{tab.1} and Eq.~(\ref{eq.4.2}).}
\centering
\begin{tabular}{c|c|c|c|c|c|c}
\hline
Channel & no cut & $N_{j,\gamma,\ell^+}$ & $\Delta \phi _{\ell m}$ & $|\vec{p}^{\ell}_T|$ & $|\vec{p}_T^{miss}|$ & $\tilde{s}_U$ \\
\hline
        SM $({\rm fb})$ & $9520.8$ & $3016.6$ & $211.7$ & $65.1$ & $40.6$ & - \\
$O_{M_2}$ $({\rm fb})$  & $6.353$ & $4.06$ & $3.51$ & $3.45$ & $3.43$ & $0.93$ \\
$O_{M_3}$ $({\rm fb})$  & $21.05$ & $13.62$ & $12.13$ & $11.95$ & $11.90$ & $2.19$ \\
$O_{M_4}$ $({\rm fb})$  & $7.39$ & $4.81$ & $4.06$ & $3.94$ & $3.92$ & $1.03$ \\
$O_{M_5}$ $({\rm fb})$  & $25.23$ & $16.73$ & $14.75$ & $14.49$ & $14.42$ & $4.05$ \\
$O_{T_5}$ $({\rm fb})$  & $2.71$ & $1.77$ & $1.28$ & $1.25$ & $1.22$ & $0.72$ \\
$O_{T_6}$ $({\rm fb})$  & $16.92$ & $11.19$ & $8.94$ & $8.36$ & $8.26$ & $3.06$ \\
$O_{T_7}$ $({\rm fb})$  & $7.47$ & $4.97$ & $3.97$ & $3.69$ & $3.65$ & $1.43$ \\
\hline
\end{tabular}
\end{table}

\subsection{\label{level4.1}Kinematic features of aQGCs}

\begin{figure}
\includegraphics[width=0.49\textwidth]{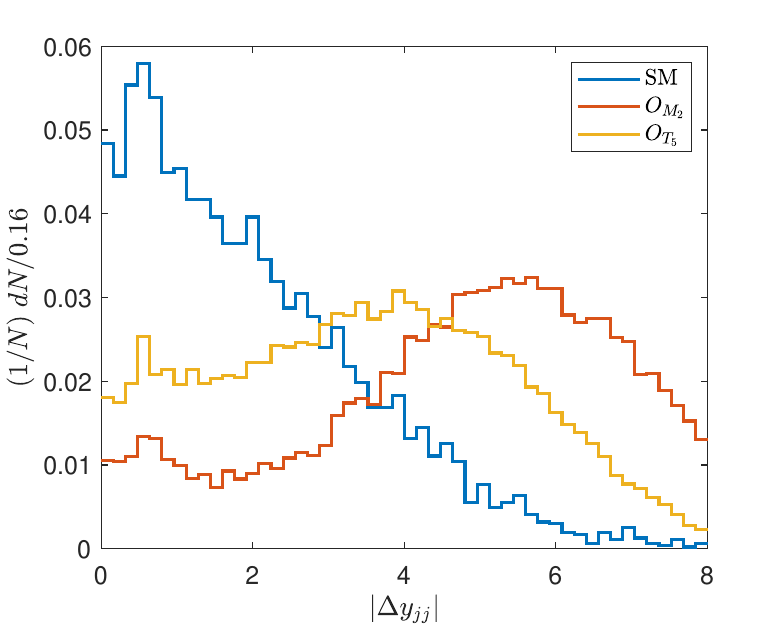}\hfill
\includegraphics[width=0.49\textwidth]{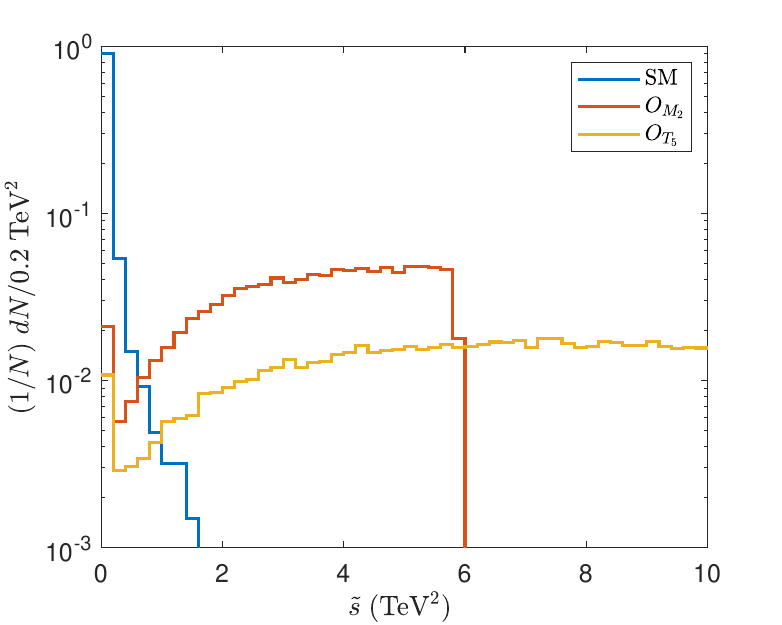}\vfill
\includegraphics[width=0.5\textwidth]{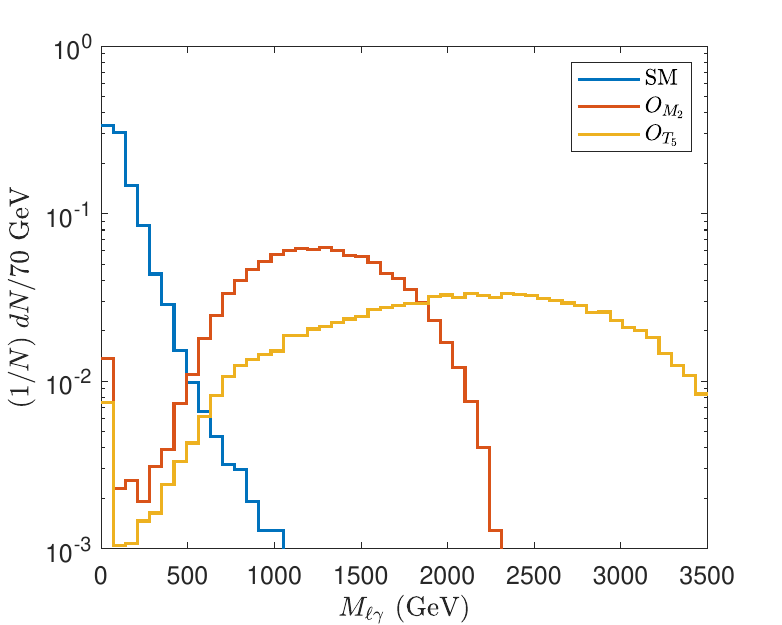}\hfill
\caption{The normalized distributions of $|\Delta y_{jj}|$, $\tilde{s}$ and $M_{\ell\gamma}$ after $\tilde{s}_U$ cut.}
\label{fig:cuts1}
\end{figure}

As introduced, in the SM, the VBS processes do not grow with $\sqrt{\hat{s}}$, which opens a window to detect aQGCs.
To focus on the VBS contributions, we investigate the efficiencies of the standard VBS/VBF cuts~\cite{VBFCut}.
The VBS/VBF cuts are designed to highlight the VBS contributions both from the SM and BSM, while can not cut off the SM VBS contributions. So they are not as efficient as other cuts designed for aQGCs only. We only impose $|\Delta y_{jj}|$ which is defined as the difference between the pseudo rapidities of the two hardest jets.
The normalized distributions of $|\Delta y_{jj}|$ are shown in Fig.~\ref{fig:cuts1}.~(a).
We find that $|\Delta y_{jj}|$ is an efficient cut for $O_{M_i}$ operators, and we select the events with $|\Delta y_{jj}| > 1.5$.

For lepton and photon, the cuts are mainly to select events with large $\hat{s}$.
The normalized distributions of $\tilde{s}$ are shown in Fig.~\ref{fig:cuts1}.(b).
We select the events with $\tilde{s}> 0.4 \;{\rm TeV}^2$.
To distinguish with the $\tilde{s}_U$ cut, $\tilde{s}$ cut in this subsection is denoted as $\tilde{s}^{\rm cut}$.

There are other sensitive observables to select large $\hat{s}$ events, such as the invariant mass of the charged lepton and photon defined as $M_{\ell \gamma}=\sqrt{(p_{\ell}+p_{\gamma})^2}$, and the angle between the photon and charged lepton, etc.
We find that, after the $\tilde{s}^{\rm cut}$ cut, the other cuts are redundant.
Take $M_{\ell \gamma}$ as an example, the normalized distributions of $M_{\ell \gamma}$ are shown in Fig.~\ref{fig:cuts1}.~(c).
As shown, $M_{\ell \gamma}$ is a very sensitive observable, and $M_{\ell \gamma} > {M^{\rm cut}_{\ell \gamma}}$ can be used as an efficient cut. However, note that after $\tilde{s}^{\rm cut}$, due to $M_{\ell \gamma}\leq \sqrt{\tilde{s}}$, one must choose a very large ${M^{\rm cut}_{\ell \gamma}}$, which is almost equivalent to a large $\tilde{s}^{\rm cut}$.

\subsection{\label{level4.2}Polarization features of aQGCs}

\begin{figure}
\begin{center}
\includegraphics[width=0.7\textwidth]{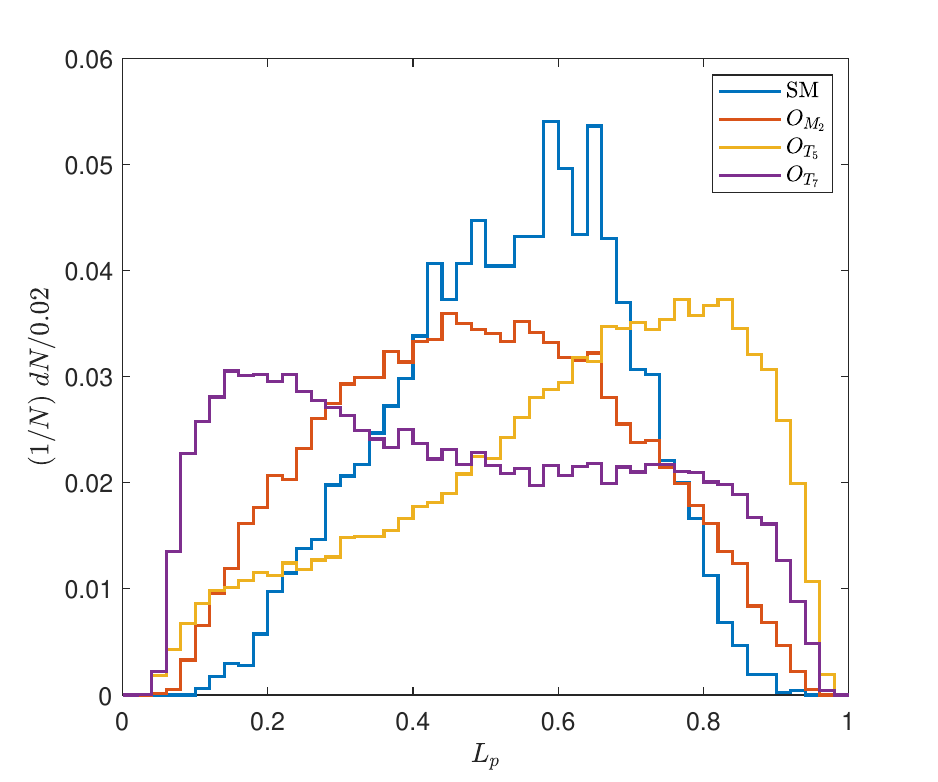}
\caption{The normalized distributions of $L_p$.}
\label{fig:polarization1}
\end{center}
\end{figure}

To improve the event select strategy, we investigate the polarization features which are less correlated with $\tilde{s}$.
One can see from Tables~\ref{tab.awaw} and \ref{tab.zwaw}, for $O_{M_i}$, the leading contributions of the signals are those with longitudinal $W^+$ bosons in the final states, while for $O_{T_i}$, both the left- and right-handed $W^+$ bosons dominate. The polarization of the $W^+$ boson can be inferred by the momentum of the charged lepton in the $W^+$ boson rest-frame, the so called helicity frame, as~\cite{wfraction}
\begin{equation}
\begin{split}
&\frac{d\sigma}{d\cos \theta^*}\propto f_L \frac{(1-\cos (\theta ^*))^2}{4}+f_R\frac{(1+\cos (\theta ^*))^2}{4} +f_0 \frac{\sin^2(\theta ^*)}{2},
\end{split}
\label{eq.4.3}
\end{equation}
where $\theta^*$ is the angle between the flight directions of $\ell^+$ and $W^+$ in the helicity frame, $f_L$, $f_R$ and $f_0=1-f_L-f_R$ are the fractions of the left-, right-handed and longitudinal polarizations, respectively.
Since the neutrinos are invisible,
it is difficult to reconstruct the momentum of the $W^+$ boson and boost the lepton to the rest frame of $W^+$ boson. However, when the transverse momentum of the $W^+$ boson is large, $\cos (\theta ^*)$ can be obtained approximately as $\cos (\theta ^*) \approx 2(L_p - 1)$ with $L_p$ defined as~\cite{LP}
\begin{equation}
\begin{split}
&L_p=\frac{\vec{p}_T^{\ell}\cdot \vec{p}_T^W}{|\vec{p}_T^W|^2},
\end{split}
\label{eq.4.4}
\end{equation}
where $\vec{p}_T^W=\vec{p}_T^{\ell}+\vec{p}_T^{miss}$.
For the signal events of $O_{M_i}$ or $O_{T_i}$ operators, the polarization fractions of $W^+$ bosons are different from those in the SM backgrounds.
We find that the polarization fractions can be categorized as four patterns, the SM pattern, the $O_{M_i}$ pattern, the $O_{T_{0,5}}$ pattern and the $O_{T_{1,2,6,7}}$ pattern.
$O_{M_2}$, $O_{T_5}$ and $O_{T_7}$ are chosen as the representations.
Neglecting the events with $L_p\notin [0, 1]$, the normalized distributions of $L_p$ after $\tilde{s}_U$ cuts are shown in Fig.~\ref{fig:polarization1}.

\begin{figure}
\subfloat[SM]{\includegraphics[width=0.48\textwidth]{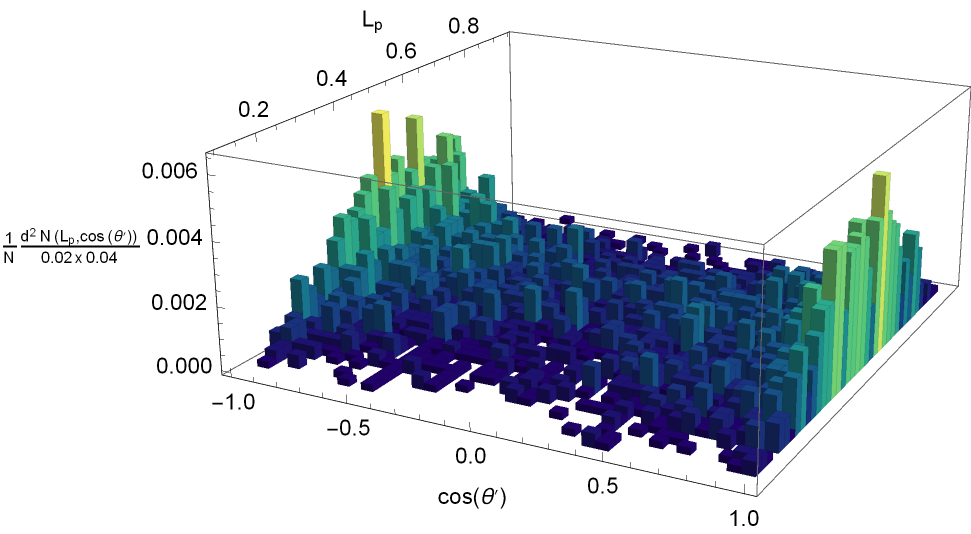}}\hfill
\subfloat[$O_{M_2}$]{\includegraphics[width=0.48\textwidth]{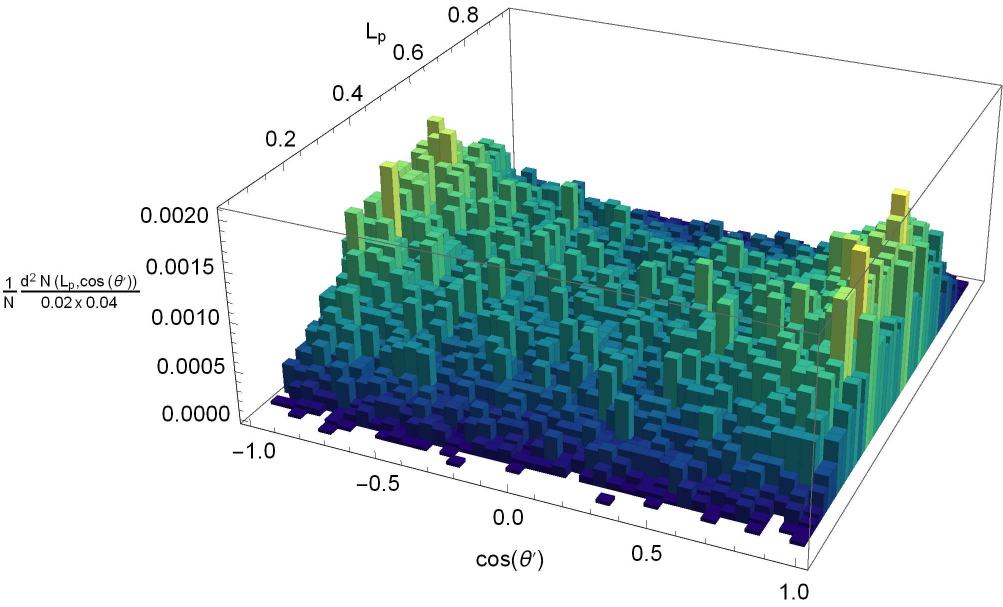}}\vfill
\subfloat[$O_{T_5}$]{\includegraphics[width=0.48\textwidth]{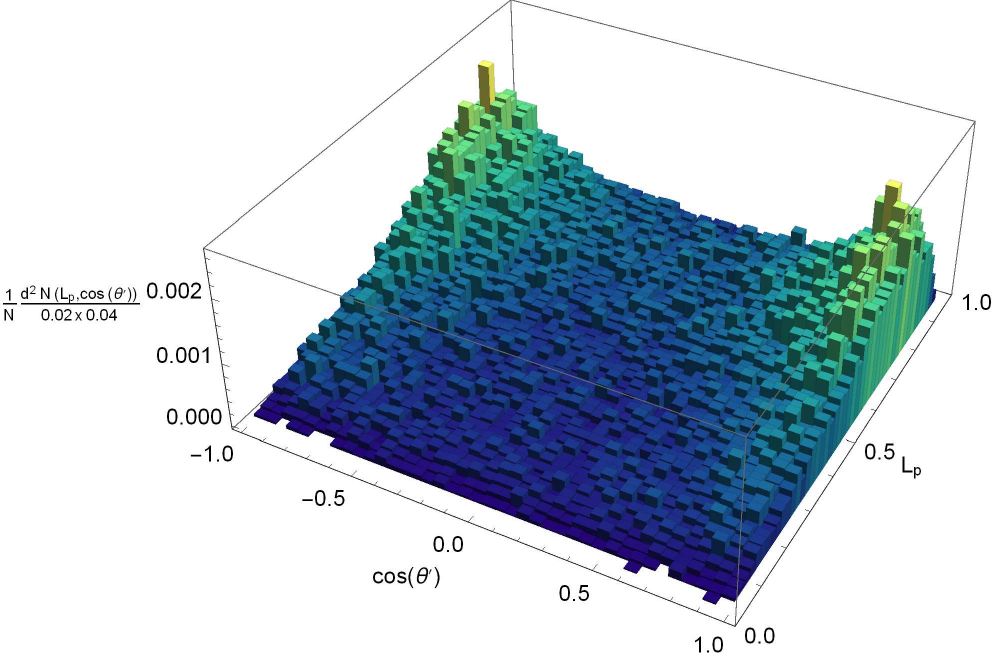}}\hfill
\subfloat[$O_{T_7}$]{\includegraphics[width=0.48\textwidth]{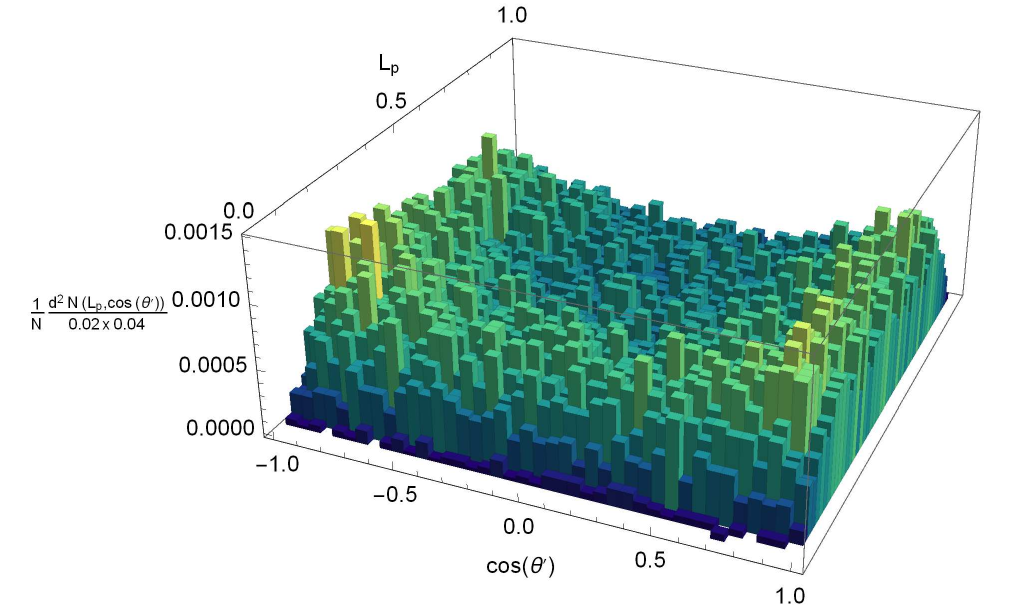}}\vfill
\caption{The normalized distributions of $L_p$ and $\cos \theta '$. Each bin corresponds to $d L_p \times d \left(\cos \theta'\right)=0.02\times 0.04$~($50\times 50$ bins).}
\label{fig:polarization2}
\end{figure}

\begin{figure}
\includegraphics[width=0.7\textwidth]{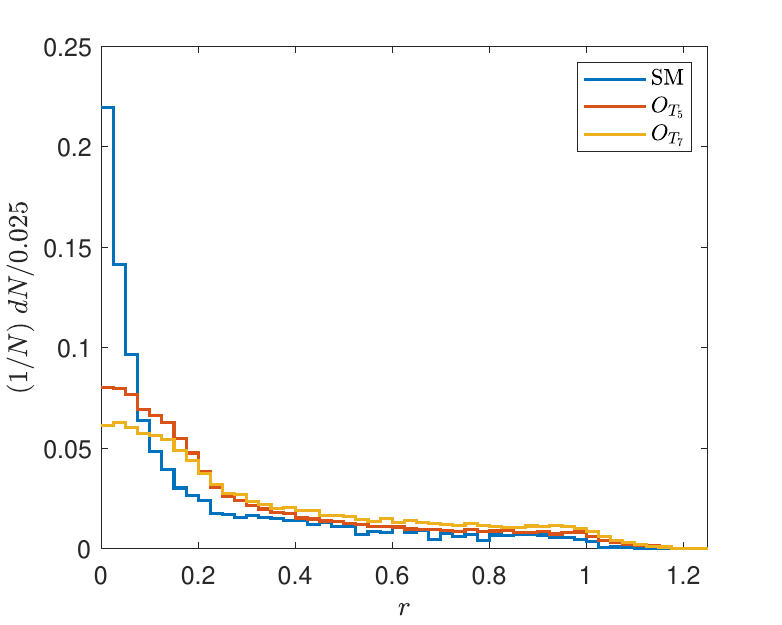}
\caption{The normalized distributions of $r$ after $\tilde{s}_U$ cut.}
\label{fig:polarization3}
\end{figure}

As presented in Tables~\ref{tab.awaw} and \ref{tab.zwaw}, the polarization of $W^+$ boson is related to $\theta$ which is the angle between the outgoing photon and the $\vec{z}$-axis of c.m. frame of the sub-process, but $\theta$ is not an observable.
Since the protons are energetic, we assume that the vector bosons in the initial states of the sub-processes carry large fractions of proton momenta, therefore the flight directions of which are close to the protons in c.m. frame. By doing so $\theta$ could be approximately estimated by the angle between outgoing photons and $\vec{z}$-axis of c.m. frame of protons, which is denoted as $\theta '$.
The correlation features between $\theta '$ and $L_p$ can be used to extract the aQGC signal events from the SM backgrounds.
The correlations of $\theta'$ and $L_p$ for the SM, and for the $O_{M_2}$, $O_{T_5}$ and $O_{T_7}$ operators are established in Fig.~\ref{fig:polarization2}.
From Fig.~\ref{fig:polarization2}, one can find that signal events of $O_{T_{5,7}}$ distribute differently from the SM backgrounds. While the distribution for the SM peaks at $|\cos(\theta ')|\approx 1$ and $L_p\approx 0.5$, the distribution for $O_{T_5}$ peaks at $|\cos(\theta ')|\approx 1$ and $L_p\approx 0$, and the distribution for $O_{T_7}$ peaks at $|\cos(\theta ')|\approx 1$ and $L_p\approx 1$. Therefore, we define
\begin{equation}
\begin{split}
&r=\left(1-\left|\cos (\theta')\right|\right)^2+\left(\frac{1}{2}-L_p\right)^2,
\end{split}
\label{eq.4.5}
\end{equation}
where $r$ is a sensitive observable can be used as a cut to discriminate the signals of the $O_{T_{5,6,7}}$ operators from the SM backgrounds.
The normalized distributions are shown in Fig.~\ref{fig:polarization3}. We select the events with $r>0.05$.

\begin{figure}
\subfloat[SM backgrounds, correlation between $M_{\ell\gamma}$ and $\tilde{s}$]{\includegraphics[width=0.48\textwidth]{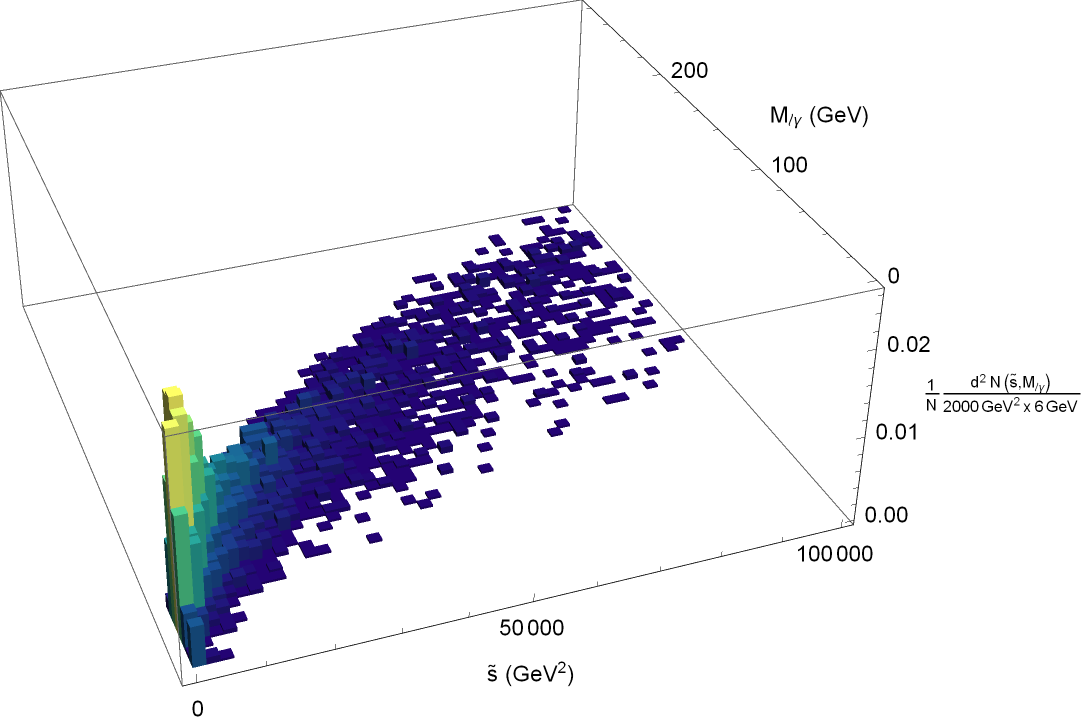}}\hfill
\subfloat[$O_{T_5}$, correlation between $M_{\ell\gamma}$ and $\tilde{s}$]{\includegraphics[width=0.48\textwidth]{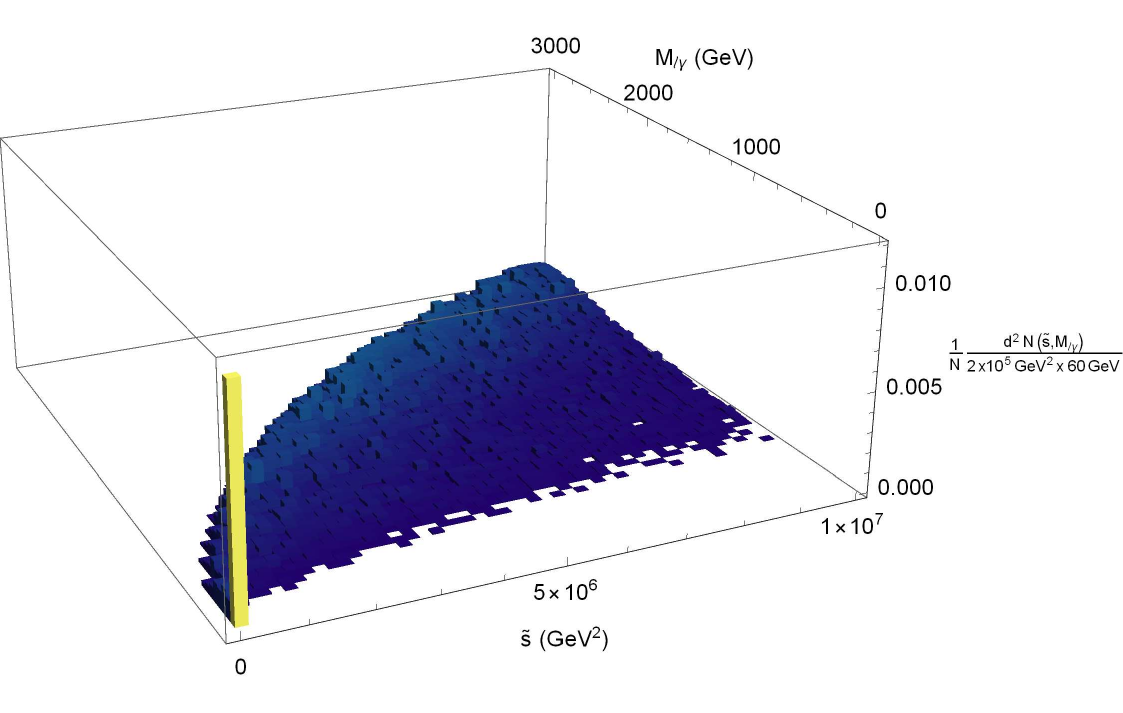}}\vfill
\subfloat[SM backgrounds, correlation between $r$ and $\tilde{s}$]{\includegraphics[width=0.48\textwidth]{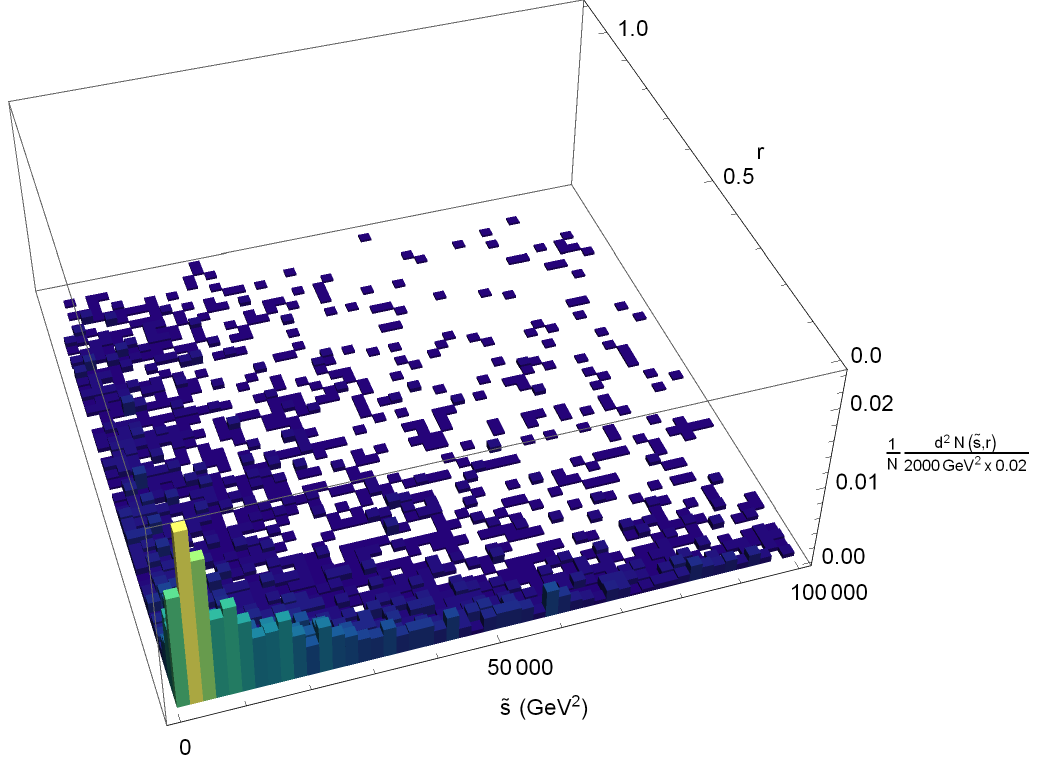}}\hfill
\subfloat[$O_{T_5}$, correlation between $r$ and $\tilde{s}$]{\includegraphics[width=0.48\textwidth]{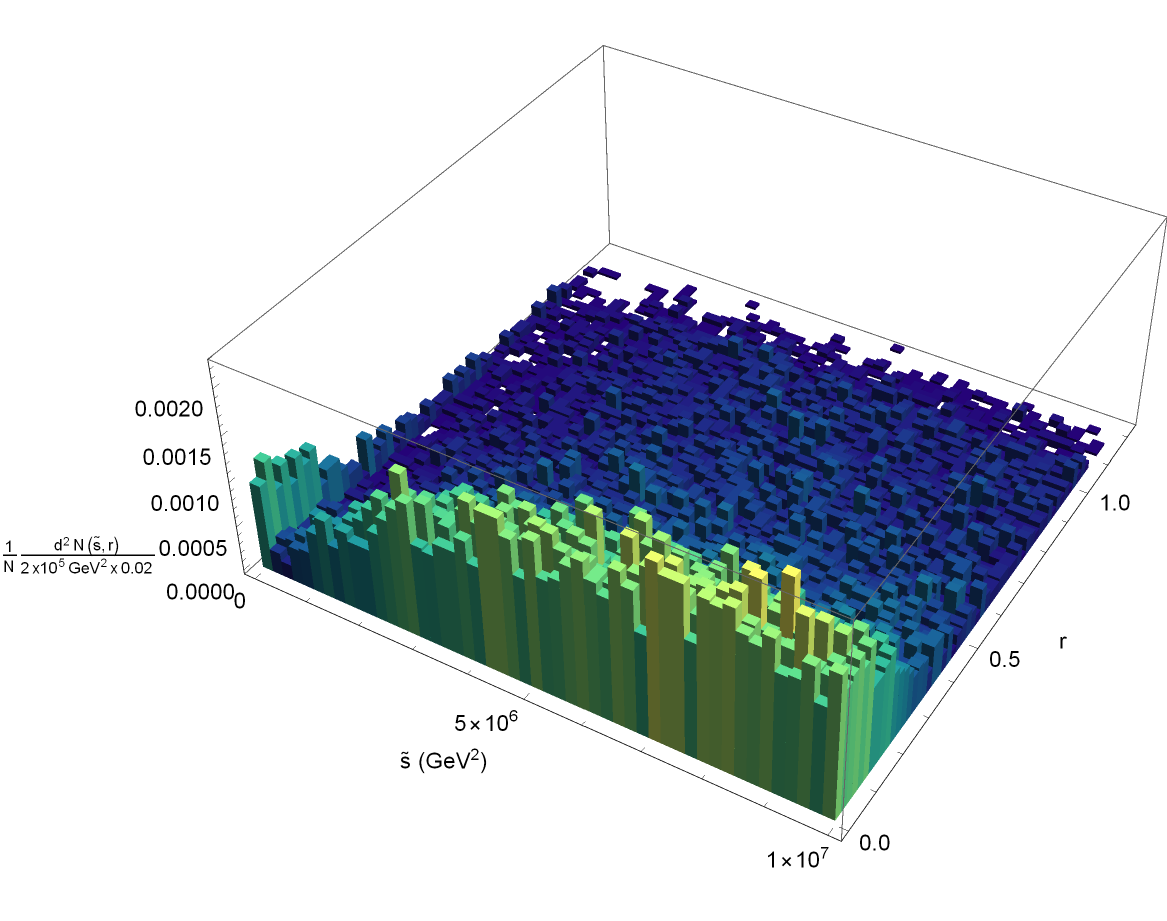}}\vfill
\caption{The correlations between $M_{\ell\gamma}$ and $\tilde{s}$ (the upper panels), $r$ and $\tilde{s}$ (the bottom panels) for $O_{T_5}$ and the SM backgrounds.}
\label{fig:mear2corr}
\end{figure}

To verify $r$ cut is not redundant, we calculate the correlation between $\tilde{s}$ and $M_{\ell \gamma}$, and compare it with the correlation between $\tilde{s}$ and $r$. Take the SM backgrounds and signal of $O_{T_5}$ as examples, the results are shown in Fig.~\ref{fig:mear2corr}. One can see that the events with small $M_{\ell \gamma}$ are almost those with small $\tilde{s}$, while it is not the case for $r$.

\subsection{\label{level4.3}Summary of the cuts}

For different operators, the kinematic and polarization features are different.
Therefore we propose to use different cuts to search for different operators, which are summarised in Table~\ref{tab.cuts}.
Note that $\tilde{s}^{\rm cut}$ in fact also cut off all the events with small $M_{\ell\gamma}$, therefore $\left|M_{\ell\gamma}-M_Z\right|>10\;{\rm GeV}$ is satisfied. The latter is used to reduce the backgrounds from $Z\to \ell\ell$ with one $\ell$ mis-tagged as a photon in the previous study of $W\gamma jj$ production~\cite{aw8TeV}, and $\tilde{s}^{\rm cut}$ has the similar effect.

\begin{table}
\caption{\label{tab.cuts}The two classes of cuts.}
\centering
\begin{tabular}{c|c}
\hline
 $O_{M_i}$ & $O_{T_{5,6,7}}$ \\
\hline
$\tilde{s}>0.4$ ${\rm TeV}^2$ & $\tilde{s}>0.4$ ${\rm TeV}^2$ \\
$|\Delta y_{jj}|>1.5$ & $0\leq L_p\leq 1$, $r>0.05$ \\
\hline
\end{tabular}
\end{table}

The results are shown in Table~\ref{tab.cutm}. The statistical error is negligible compared with the systematic error, therefore is not presented. The large SM backgrounds can be effectively reduced by our selection strategy.

\begin{table}
\caption{\label{tab.cutm}The cross sections (fb) of signals and the SM backgrounds after $\tilde{s}^{\rm cut}$, $|\Delta y_{jj}|$ and $r$ cuts.
The column `after $\tilde{s}_U$' is as same as the last column of Table~\ref{tab.unitaritybound}.}
\centering
\begin{tabular}{c|c|c|c}
\hline
Channel & after $\tilde{s}_U$ & after $\tilde{s}^{\rm cut}$ & $|\Delta y_{jj}|$ or $r$ \\
\hline
       SM   & $40.6$ & $1.70$ & $0.93^{+0.23}_{-0.17}$ ($\Delta y_{jj}$) \\
                        &         &       & $1.05^{+0.26}_{-0.19}$ ($r$) \\
$O_{M_2}$   & $0.93$ & $0.91$ & $0.82^{+0.20}_{-0.15}$ \\
$O_{M_3}$   & $2.19$ & $2.11$ & $1.90^{+0.48}_{-0.35}$ \\
$O_{M_4}$   & $1.03$ & $1.01$ & $0.91^{+0.23}_{-0.16}$ \\
$O_{M_5}$   & $4.05$ & $3.94$ & $3.55^{+0.89}_{-0.64}$ \\
$O_{T_5}$   & $0.72$ & $0.71$ & $0.60^{+0.15}_{-0.11}$  \\
$O_{T_6}$   & $3.06$ & $3.01$ & $2.69^{+0.62}_{-0.48}$  \\
$O_{T_7}$   & $1.43$ & $1.40$ & $1.12^{+0.28}_{-0.20}$  \\
\hline
\end{tabular}
\end{table}

\section{\label{level5}Cross sections and statistical significances}

To investigate the signals of aQGCs, one should investigate how the cross section is modified by adding dimension-8 operators to the SM Lagrangian, the effect of interference is also included. In this section, we investigate the process $pp\to \ell^+ \nu \gamma j j$ with all Feynman diagrams including non-VBS aQGC diagrams, such as Fig.~\ref{fig:typicalSignal}.~(b), and with all possible interference effects.

\begin{figure}
\subfloat[without unitarity bounds]{\includegraphics[width=0.49\textwidth]{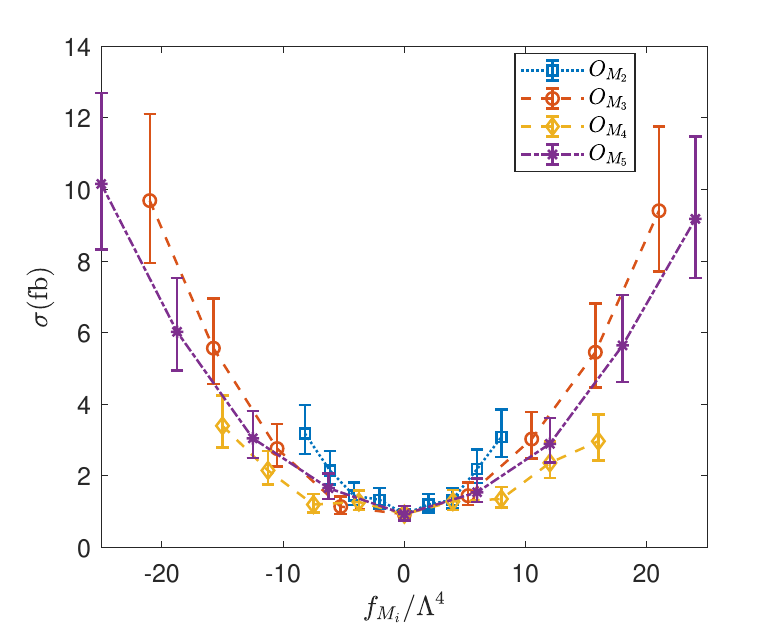}}\hfill
\subfloat[with unitarity bounds]{\includegraphics[width=0.49\textwidth]{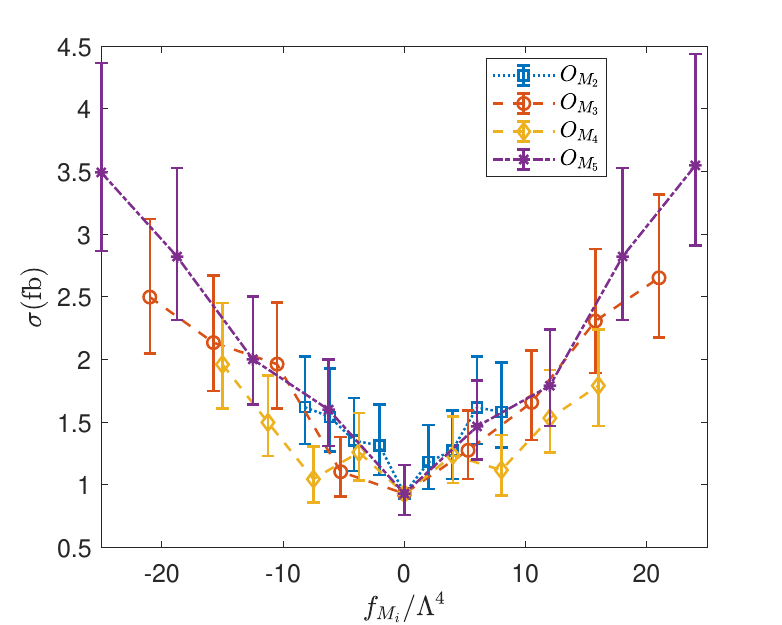}}\vfill
\caption{The cross sections as functions of $f_{M_i/\Lambda^4}$ with and without unitarity bounds.}
\label{fig:fitting}
\end{figure}

\begin{figure}
\subfloat[without unitarity bounds]{\includegraphics[width=0.49\textwidth]{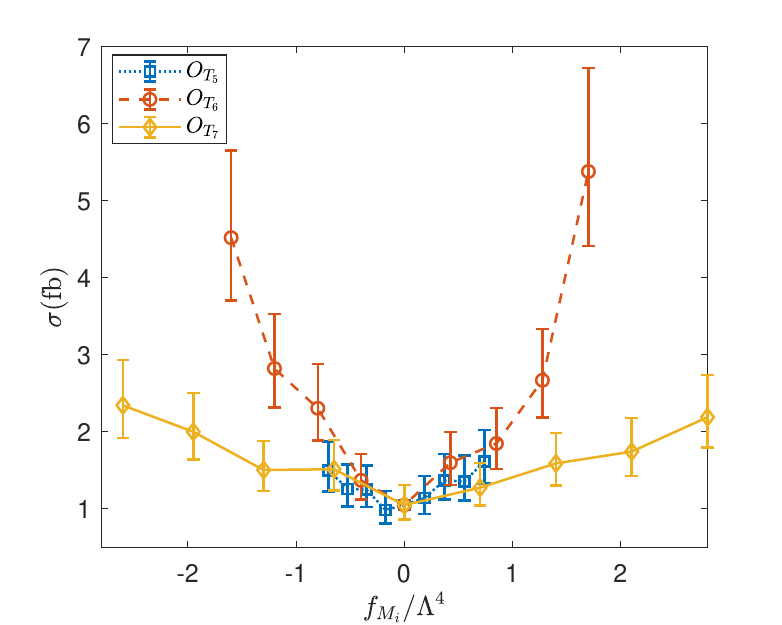}}\hfill
\subfloat[with unitarity bounds]{\includegraphics[width=0.49\textwidth]{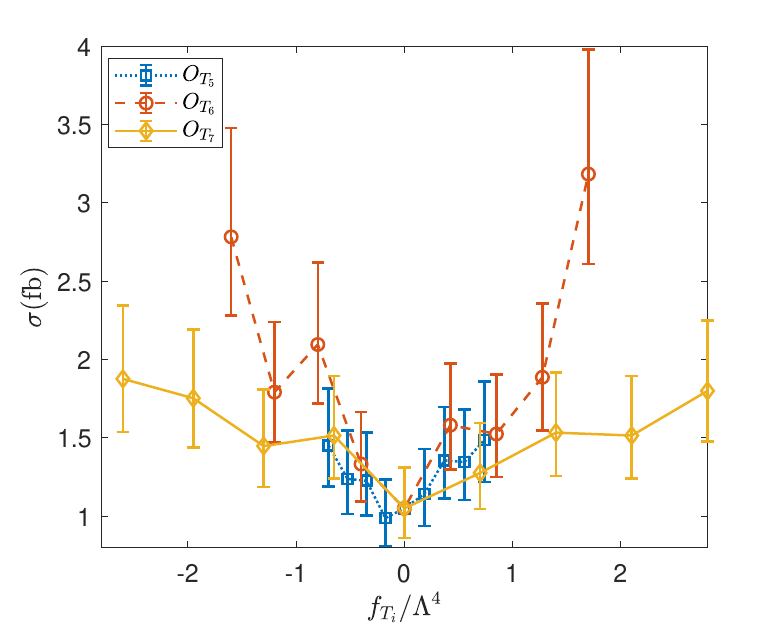}}\vfill
\caption{The cross sections as functions of $f_{T_i/\Lambda^4}$ with and without unitarity bounds.}
\label{fig:fitting2}
\end{figure}

To investigate the parameter space, we generate events with one operator at a time. The unitarity bounds are set as $\tilde{s}_U$ cuts which depend on $f_{M_i}/\Lambda ^4$ and $f_{T_i}/\Lambda ^4$ used to generate the events. The cross sections as functions of $f_{M_i}/\Lambda ^4$ and $f_{T_i}/\Lambda ^4$ are shown in Figs.~\ref{fig:fitting} and \ref{fig:fitting2}. The results with and without the unitarity bounds are both presented, one can find from Figs.~\ref{fig:fitting} and \ref{fig:fitting2} that, without the unitarity bounds, the cross sections are approximately bilinear functions of $f_{M_i}/\Lambda ^4$ and $f_{T_i}/\Lambda ^4$. However, the unitarity bounds greatly suppress the signals, and the resulting cross sections are no longer bilinear functions. One can also find from Figs.~\ref{fig:fitting} and \ref{fig:fitting2} that the $ W\gamma jj$ production is more sensitive to the $O_{M_{3,5}}$ and $O_{T_{6,7}}$ operators.

The constraints on the operator coefficients can be estimated with the help of statistical significance defined as $\mathcal{S}_{stat}\equiv N_S/\sqrt{N_S+N_B}$, where $N_S$ is the number of signal events, and $N_B$ is the number of the background events. The total luminosity $\mathcal{L}$ at $13\;{\rm TeV}$ for the years 2016, 2017 and 2018 sum up to about $\mathcal{L}\approx 137.1\;{\rm fb^{-1}}$~\cite{lumino}. For each $f_X/\Lambda ^4$ used to generate the events, the $\tilde{s}_U$ cut can be set accordingly, afterwards $\mathcal{S}_{stat}$ at $137.1\;{\rm fb^{-1}}$ and $13\;{\rm TeV}$ can be obtained. The constraints are set by the lowest positive $f_X/\Lambda ^4$ and greatest negative $f_X/\Lambda ^4$ with $\mathcal{S}_{stat}$ larger than the required statistical significance. The constraints on the coefficients at current luminosity are shown in Table~\ref{tab.coefficients}.
By comparing the constraints from 13 TeV CMS experiments in Table~\ref{tab.1} with the ones in Table~\ref{tab.coefficients}, one can find that, even with the unitarity bounds suppressing the signals, using our efficient event selection strategy, the allowed parameter space can still be reduced significantly.

\begin{table}
\caption{\label{tab.coefficients}The constraints on the operators at LHC with $\mathcal{L}=137.1\; {\rm fb^{-1}}$.}
\begin{tabular}{c|c||c|c}
\hline
Coefficients & $\mathcal{S}_{stat}> 2$ & Coefficients & $\mathcal{S}_{stat}> 2$\\
\hline
$f_{M_2}/\Lambda ^4$ & $[-2.05, 2.0]$ & $f_{T_5}/\Lambda ^4$ & $[-0.525, 0.37]$\\
$f_{M_3}/\Lambda ^4$ & $[-10.5, 5.25]$ & $f_{T_6}/\Lambda ^4$ & $[-0.4, 0.425]$\\
$f_{M_4}/\Lambda ^4$ & $[-11.25, 4.0]$ & $f_{T_7}/\Lambda ^4$ & $[-0.65, 0.7]$\\
$f_{M_5}/\Lambda ^4$ & $[-6.25, 6.0]$ & & \\
\hline
\end{tabular}
\end{table}

\section{\label{level6}Summary}

The accurate measurement of the VBS processes at the LHC is very important for the understanding of the SM and search of BSM.
In recent years, the VBS processes drew a lot of attention, and have been studied extensively.
To investigate the signals of BSM, a model independent approach known as the SMEFT is frequently used, and the effects of BSM show up as higher dimensional operators.
The VBS processes can be used to probe dimension-8 anomalous quartic gauge-boson operators.
In this paper, we focus on the effects of aQGCs in the process $pp\to W \gamma jj$.
The operators concerned are summarized, and the corresponding vertices are obtained.

An important issue of the SMEFT is its validity.
We study the validity of the SMEFT by using the partial-wave unitarity bound, which sets an upper bound on $\hat{s}^2 |f_X|$ where $f_X$ is the coefficient of the operator $O_X$.
In other words, there exist a maximum $\hat{s}$ for a fixed coefficient in the sense of unitarity.
We discard all the events with $\hat{s}$ larger than the maximally allowed one, then the results obtained via the SMEFT are guaranteed to respect unitarity.
For this purpose, we find an observable which can approximate $\hat{s}$ very well denoted as $\tilde{s}$, based on which the unitarity bounds are applied.
Due to the fact that there are massive $W^+$ or/and $Z$ bosons in the initial state of the sub-process, and that the massive particle emitting from a proton can carry a large fraction of the proton momentum, the c.m. energy of the sub-process is found to be at the same order as the c.m. energy of corresponding process.
As a consequence, at large c.m. energy, the unitarity bounds are very strict, and the cuts can greatly reduce the signals.

To study the discovery potential of aQGCs, we investigate the kinematic features of the signals induced by aQGCs, and find that $\tilde{s}$ serves as a very efficient cut to highlight the signals.
We also find that other cuts to cut off the events with small $\hat{s}$ are redundant. To find other sensitive observables less correlated with $\hat{s}$, we investigate the polarization features of the signals.
The polarization features of $O_{T_i}$ operators are found to be very different from the SM backgrounds.
We find a sensitive observable $r$ to select the signal events of $O_{T_i}$ operators.
Although the signals of aQGCs are highly  suppressed by unitarity bounds, the constraints on the coefficients for $O_{M_{2,3,4,5}}$ and $O_{T_{5,6,7}}$ operators can still be tightened significantly with current luminosity at 13 TeV LHC.

\section*{ACKNOWLEDGMENT}

\noindent
We thank Jian Wang and Cen Zhang for useful discussion. This work was supported in part by the National Natural Science Foundation of China under Grants No.11905093, 11875157 and 11947402; and by the Doctoral Start-up Foundation of Liaoning Province No.2019-BS-154.

\end{document}